%% file: 2023-arxiv-fp-explaignn.tex
\documentclass[sigconf,natbib=true,anonymous=false]{acmart}

\AtBeginDocument{%
  \providecommand\BibTeX{{%
    \normalfont B\kern-0.5em{\scshape i\kern-0.25em b}\kern-0.8em\TeX}}}

\usepackage{latexsym}
\usepackage{graphicx}
\graphicspath{{./images/}}
\usepackage{booktabs} % for formal tables
\usepackage{color}  % for coloring text
\usepackage{amsmath}  % for aligning equations
\usepackage{subcaption}
\usepackage{caption}
\usepackage{tikz}
\usepackage{colortbl} % for color in tables
\usepackage{framed}
\usepackage{multirow}
\usepackage{multicol}
\usepackage{hyperref}
\usepackage{url}
\usepackage{balance}
\usepackage{verbatim}
\usepackage{cancel}
\usepackage{xspace} % for correcting space after macro commands
\usepackage[ruled,vlined]{algorithm2e}
\usepackage{bbold} % for writing mathbb{1}
\usepackage{balance}
\usepackage{stmaryrd}
\usepackage{enumitem}
\usepackage{array} % package for hiding a column
\usepackage{pifont} % for ticks and crosses 

\newcommand{\cmark}{\ding{51}}%
\newcommand{\xmark}{\ding{55}}%

\newcommand{\struct}[1]{\texttt{\small #1}}
\newcommand{\utterance}[1]{\textit{#1}}
\newcommand{\phrase}[1]{\textit{``#1''}}

\newcommand{\drop}{\dag\xspace}

\newenvironment{Snugshade}[1][236,236,236]{
    \setlength{\itemsep}{0pt}
     \setlength{\parsep}{0pt}
     \setlength{\topsep}{0pt}
     \setlength{\partopsep}{0pt}
     \setlength{\leftmargin}{1.5em}
     \setlength{\labelwidth}{0em}
     \setlength{\labelsep}{0em} 
    \setlength{\parskip}{0pt}
    \definecolor{shadecolor}{RGB}{#1}%
    \begin{snugshade}
}{%
    \end{snugshade}%
}

\newcommand{\explaignn}{\textsc{Explaignn}\xspace}

\newcommand{\convinse}{\textsc{Convinse}\xspace}
\newcommand{\convmix}{\textsc{ConvMix}\xspace}
\newcommand{\clocq}{\textsc{Clocq}\xspace}
\newcommand{\convquestions}{\textsc{ConvQuestions}\xspace}

\newcommand {\itemslist}[1]{$\langle$\struct{#1}$\rangle$}

\copyrightyear{2023} 
\acmYear{2023} 
\setcopyright{rightsretained} 
\acmConference[SIGIR '23]{Proceedings of the 46th International ACM SIGIR Conference on Research and Development in Information Retrieval}{July 23--27, 2023}{Taipei, Taiwan}
\acmBooktitle{Proceedings of the 46th International ACM SIGIR Conference on Research and Development in Information Retrieval (SIGIR '23), July 23--27, 2023, Taipei, Taiwan}

\begin{document}

% to remove default header and footer
% \fancyhead{}
% \fancyfoot{}
% \pagenumbering{gobble}

\title{Explainable Conversational Question Answering over Heterogeneous Sources via Iterative Graph Neural Networks}
% \title{Explainable Conversational Question Answering over Heterogeneous Sources via Graph Neural Networks}

\author{Philipp Christmann}
%\orcid{1234-5678-9012}
\affiliation{%
  \institution{Max Planck Institute for Informatics\\Saarland Informatics Campus}
  \streetaddress{Saarland Informatics Campus}
  \city{Saarbruecken}
 \country{Germany}
%  \postcode{43017-6221}
}
\email{pchristm@mpi-inf.mpg.de}

\author{Rishiraj Saha Roy}
%\orcid{1234-5678-9012}
\affiliation{%
  \institution{Max Planck Institute for Informatics\\Saarland Informatics Campus}
  \streetaddress{Saarland Informatics Campus}
  \city{Saarbruecken}
    \country{Germany}
  %  \postcode{43017-6221}
}
\email{rishiraj@mpi-inf.mpg.de}

\author{Gerhard Weikum}
%\authornote{Dr.~Trovato insisted his name be first.}
%\orcid{1234-5678-9012}
\affiliation{%
  \institution{Max Planck Institute for Informatics\\Saarland Informatics Campus}
  \streetaddress{Saarland Informatics Campus}
  \city{Saarbruecken}
    \country{Germany}
  %  \postcode{43017-6221}
}
\email{weikum@mpi-inf.mpg.de}

% The default list of authors is too long for headers.
\renewcommand{\shortauthors}{Christmann et al.}

\newcommand{\squishlist}{
    \begin{list}{$\bullet$}{ 
        \setlength{\itemsep}{0pt}
        \setlength{\parsep}{1pt}
        \setlength{\topsep}{1pt}
        \setlength{\partopsep}{0pt}
        \setlength{\leftmargin}{1.5em}
        \setlength{\labelwidth}{1em}
        \setlength{\labelsep}{0.5em} 
    } 
}

\newcommand{\squishend}{
  \end{list}  }
  
\newcommand{\GW}[1]{{\color{blue}{GW: #1}} }
\newcommand{\RSR}[1]{{\color{red}{RSR: #1}} }
\newcommand{\PC}[1]{{\color{orange}{PC: #1}} }

\newcommand{\myparagraph}[1]{\noindent \textbf{#1}.}

\setcounter{secnumdepth}{4}

\input{sections/00-abstract}


% \keywords{Question answering, Conversations, Heterogeneous sources}
% \keywords{Question Answering, Explainability, Graph Neural Networks}

\maketitle
\input{sections/01-intro}

\input{sections/02-concepts}

\input{sections/03-method}
\input{sections/04-experiments.tex}

\input{sections/05-results}
% 2 pages
\input{sections/06-explainability.tex}

% 1 page
\input{sections/07-related-work}

\input{sections/08-conclusion}
\balance

\bibliographystyle{ACM-Reference-Format}
\bibliography{explaignn}

\end{document}

%% file: sections/00-abstract.tex
% !TEX root = ../2023-sigir-fp-explaignn.tex
\begin{abstract}
In conversational question answering,
% (ConvQA),
users express their information needs through a series of utterances with incomplete context.
% and ad hoc style.
% Existing
Typical ConvQA methods
% typically
rely on a single
% information
source
%, like a
% curated
(a knowledge base (KB), {\em or} a text
corpus,
% collection,
{\em or} a set of
% web
tables),
thus being unable to benefit from increased answer coverage and redundancy of multiple sources.
% thereby reducing the overall answer recall.
% Further, none % of them
% provide explanations that support the answer derivation.
% process.
% We propose \explaignn: a
Our method \explaignn overcomes these limitations by integrating information from a mixture of sources with user-comprehensible explanations for answers.
% Our technique
It constructs a heterogeneous graph from entities and evidence snippets retrieved from a KB, a text corpus, web tables, {\em and} infoboxes. This large graph is then iteratively reduced via graph neural networks that incorporate question-level attention, until the best answers and their explanations are distilled out.
% Comprehensive
Experiments show that \explaignn improves
% answering
performance over state-of-the-art
% ConvQA
baselines. % , and a 
% crowdsourced
A user study demonstrates that derived answers % derived
% by the proposed framework
are understandable by end users.
\end{abstract}

%% file: sections/01-intro.tex
\begin{figure*} [t]
     \includegraphics[width=\textwidth]{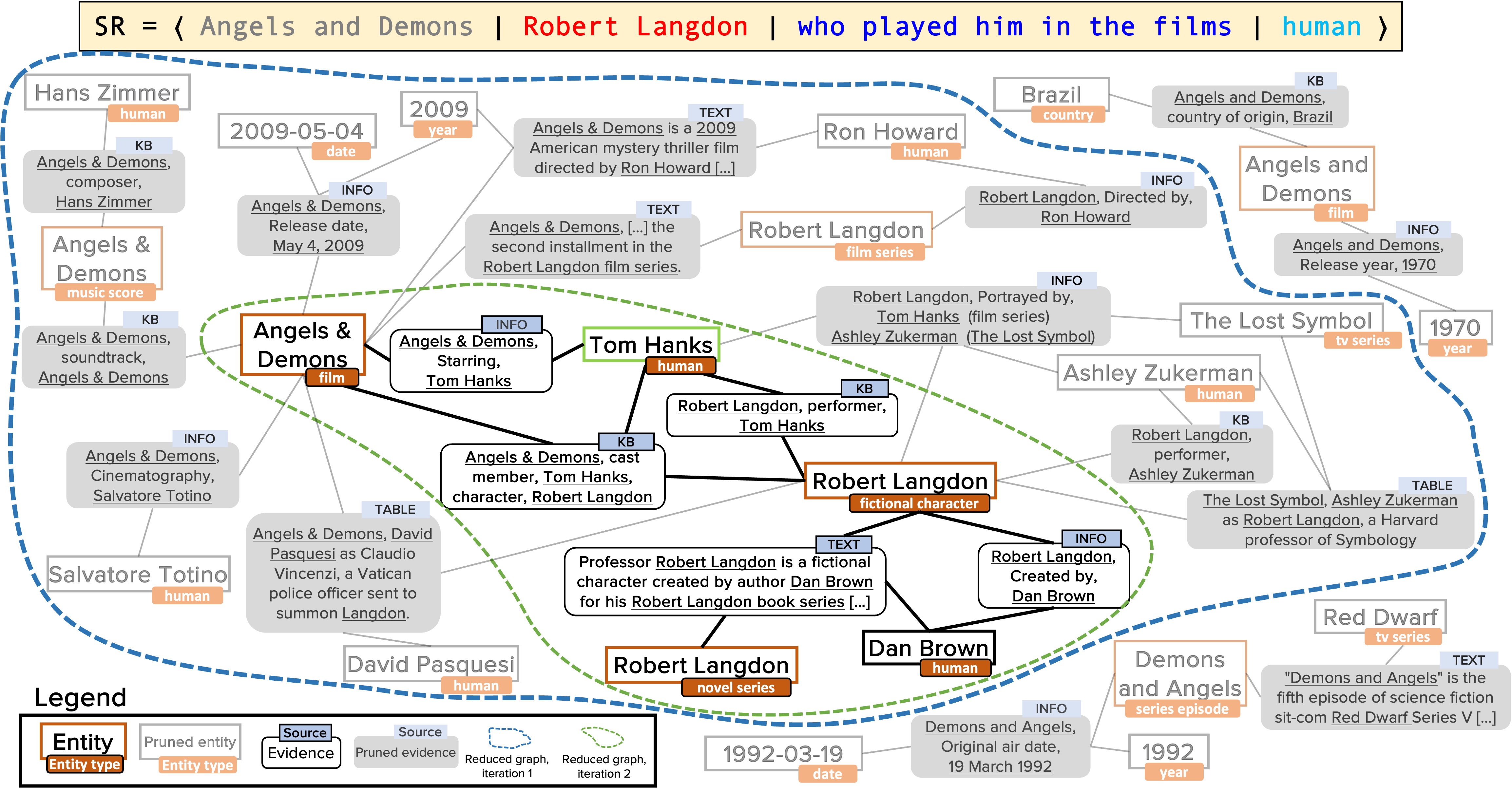}
     \vspace*{-0.4cm}
     \caption{Toy heterogeneous graph for answering $q^3$, showing two pruning iterations. The graph is iteratively reduced by GNN inference
     %in size to focus on the relevant information. 
     to identify
     % the answer
     the key evidences. 
     The subgraph surrounded by the blue dotted line is the result of the first iteration, while the green 
     % dotted
     line indicates the graph after the second.
     % iteration.
     From this smaller subgraph, the final answer (\struct{Tom Hanks}) is inferred.}
     \label{fig:graph}
     \vspace*{-0.3cm}
\end{figure*}

% !TEX root = ../2023-sigir-fp-explaignn.tex
\section{Introduction}
\label{sec:intro}

% motivation of problem (convqa over heterogeneous sources)
\myparagraph{Motivation}
In conversational question answering (ConvQA), 
users issue a
sequence of 
% series of 
%knowledge-centric 
questions, and the 
%The
ConvQA 
system
% is then supposed to derive crisp answers for these information needs 
% is supposed to
computes crisp answers~\cite{saharoy2022question, reddy2019coqa, qu2020open}.
%
%Deriving the answers 
The main challenge in ConvQA systems is that
inferring answers
requires understanding the 
current
% ongoing
% conversational
context,
since incomplete, ungrammatical and informal \textit{follow-up questions} %important information is often kept implicit in .
% often
make sense only when considering the
% entire
conversation
history
so far.
%Also, follow-up questions can be 
% Such questions are often 
% incomplete, informal or ungrammatical, and understandable only 
% with proper context.
% This poses major challenges for ConvQA systems.
% in a context-sensitive manner.
%ad-hoc manner,
%with ungrammatical formulations, posing key challenges for ConvQA systems.

%Existing work 
% Prior works
Existing ConvQA models
% so far
mostly focused on 
%answering conversational questions 
using
either (i) a curated knowledge base (KB)~\cite{kaiser2021reinforcement,shen2019multi,kacupaj2022contrastive,marion2021structured,lan2021modeling,ke2022knowledge}, or (ii) a text corpus~\cite{chengraphflow,huang2018flowqa,qiu2021reinforced,qu2019attentive,qu2020open}, or (iii) a set of web tables~\cite{mueller2019answering,iyyer2017search}
as % information
source to compute answers.
These methods 
% are limited to using a sing
are not geared for tapping into multiple sources jointly, which is often crucial
% for recall 
% precision and recall discussed later
as one source could compensate for gaps in others.
%for answering different questions within a conversation:
%all questions have to be answered using the same information source.
Consider the
% following example
conversation:
% \begin{Snugshade}
% 	$q^1$: \utterance{name of Argentinas coach in the WC 2022?}
	   
%     \indent\indent $a^1$: \textit{Lionel Sebastián Scaloni}		
	
% 	$q^2$: \utterance{who assisted the second goal in the final?}	
	
% 	\indent\indent $a^2$: \textit{Alexis Mac Allister}	
	
% 	$q^3$: \utterance{which club does he play for?}
	
% 	\indent\indent $a^3$: \textit{Brighton \& Hove Albion}

% 	$q^4$: \utterance{when was he born?}	
	
% 	\indent\indent $a^4$: \textit{24 December 1998}	
	
% 	$q^5$: \utterance{who received the man of the match award?}
	
% 	\indent\indent $a^5:$ \textit{Lionel Messi}

% 	$q^6$: \utterance{whose penalty was saved by the goalkeeper?}	
% 	\indent\indent $a^6$: \textit{Kingsley Coman}
% \end{Snugshade}

\begin{Snugshade}
	$q^1$: \utterance{Who wrote the book Angels and Demons?}\\
    \indent\indent $a^1$: \struct{Dan Brown}		

	$q^2$: \utterance{the main character in his books?}\\
	\indent\indent $a^2$: \struct{Robert Langdon}

 	$q^3$: \utterance{who played him in the films?}\\
	\indent\indent $a^3$: \struct{Tom Hanks}

    $q^4$: \utterance{to which headquarters was robert flown in the book?}\\
    \indent\indent $a^4$: \struct{CERN}

    % $q^4$: \utterance{To which european research center is Robert flown in the book?}\\
	% \indent\indent $a^4$: \textit{CERN}

 %    $q^5$: \utterance{and the city was?}\\
	% \indent\indent $a^5$: \struct{Meyrin}

    $q^5$: \utterance{how long is the novel?}\\
	\indent\indent $a^5$: \struct{768 pages}

    $q^6$: \utterance{what about the movie?}\\
	\indent\indent $a^6$: \struct{2 h 18 min}
\end{Snugshade}
\noindent Some 
of these
questions
% in this conversation
can be conveniently answered using 
a KB ($q^1$, $q^3$),
tables ($q^5$, $q^6$),
or
infoboxes ($q^1$, $q^3$)
as they ask about salient attributes of entities,
and some via text sources ($q^2$, $q^3$, $q^4$)
as they are more likely to be contained in book contents and discussion.
However, none of these individual sources 
% is designed for representing the knowledge
% required
represents the whole information required
to answer \textit{all questions} of this conversation.
% Further, some questions can be answered from several information sources (e.g. $q^1$, $q^2$).

Recently, there has been preliminary work on ConvQA over
a mixture of input
% heterogeneous 
% Recent works on ConvQA over heterogeneous 
sources~\cite{christmann2022conversational, deng2022pacific}.
% have proposed graph-based techniques, including graph neural networks (GNNs),
% to jointly tap into all of the above sources.
% with the goal of integrating multiple information sources to boost answer coverage and redundancy.
%When integrating multiple information sources, 
%% it is obvious that
%% there is a higher chance of obtaining relevant information for answering a question,
%the \textit{answer coverage} is boosted.
%Also,
%% the same
%relevant information can be present in different information sources simultaneously, potentially represented in different ways.
%This \textit{answer redundancy} can help the ConvQA system to pin-point the correct answer.
This improves
the recall for the QA system
with higher
% (equivalently, the 
% coverage of answerability
\textit{answer coverage}, and the partial \textit{answer redundancy} across sources helps improve precision.

% \vspace*{0.2cm}
% limitations of state-of-the-art
\myparagraph{Limitations of state-of-the-art methods}
% However, 
% The
% best
Existing
methods for ConvQA over heterogeneous sources rely on neural sequence-to-sequence models to compute answers~\cite{christmann2022conversational, deng2022pacific}.
%Such models have been shown to achieve high metrics on a variety of tasks and datasets~\cite{raffel2020exploring, devlin2019bert}.
%% However, it is typically impossible for users to verify the correctness of the provided answer.
However, this has two significant limitations:
(i) sequence-to-sequence models are \textit{not explainable}, as they only generate
% single
strings as outputs, making it infeasible for users to decide whether to trust the answer;
% which negatively affects the \textit{explainability};
% of ConvQA systems.
%
%This is similar as for conversational agents based on 
%language model
%prompting~\cite{brown2020language, radfordlanguage}, like ChatGPT,
%and negatively affects the \textit{explainability} of ConvQA systems. 
% (e.g.  ChatGPT).
%GW: for IR, I don't think this digression is needed
%
(ii) sequence-to-sequence models require inputs to be cast into token sequences first.
% For example, in the answering stage of the pipeline proposed in~\cite{christmann2022conversational}, all evidences obtained from heterogeneous sources
% are linearized into a textual sequence, 
% %before they are given 
% as input to an answer generator~\cite{izacard2021leveraging}.
% generative model
% Fusion-in-Decoder model~\cite{izacard2021leveraging}.
%This means that all structural information among evidences is lost.
This loses insightful information on relationships between evidences~\cite{yu2022kg}.
% (e.g. when evidences share entities).
% when evidences are retrieved from
% the same web page, or point towards the same answer candidate,
% there is often some inherent connection between their content.
% are often retrieved from the same web page, or mention the same answer candidate.
Such inter-evidence connections can be helpful in separating relevant information from noise.
% , but cannot be leveraged by typical sequence-to-sequence models.
% This loses potentially important information embedded in the original connections
% graph-based pieces of evidence.
% \GW{the next two sentences are not clear to me: what is the message exactly? need to re-word this into a single crisp sentence! (or leave it at this point (abstractly, without example)!}

% Also, sequence-to-sequence models can only process a limited amount of data
% i.e. the existing lacks explainability.
% Also, the assumption that the relevant information (i.e. the relevant text document and table) is given~\cite{deng2022pacific},
% does not hold, if one aims to operate the ConvQA system in an open-domain manner.

% proposed method in a nutshell
\myparagraph{Approach}
We introduce \explaignn\footnote{Code, data, and demo at \url{https://explaignn.mpi-inf.mpg.de}.} 
(\underline{EXPLA}inable Conversational Question Answering over Heterogeneous Graphs via \underline{I}terative \underline{G}raph \underline{N}eural \underline{N}etworks),
a
flexible
pipeline 
that can be configured for 
optimizing \textit{performance, efficiency, and explainability}
% new
% efficient
% method
% that
% aims to
% enhances both the answering performance and the explainability of
for ConvQA systems over heterogeneous sources.
% The 
%approach consists of three main stages:
% method
The proposed method operates in three stages:
% \GW{itemization is important to highlight his main message}
\squishlist
\item[(i)] Derivation of a
self-contained
% intent-capturing
{\em structured representation (SR)} of the user's information need (or intent) from the potentially incomplete input utterance and the conversational context, making the % central
entities, relation, and expected answer type explicit.
\item[(ii)] Retrieval of relevant evidences and answer candidates
% cues are retrieved
from heterogeneous information sources: a curated KB, a text corpus, a collection of web tables, and infoboxes.
\item[(iii)] Construction of 
%next, a heterogeneous answering graph is constructed from 
% the set of relevant
%these evidences, and 
a graph
% is constructed
from these evidences,
% these evidences are leveraged to construct a graph,
as the basis for applying graph neural networks (GNNs).
% to overcome the outlined limitations of prior works.
The GNNs are iteratively
applied
% through their inference,
for computing the best answers and supporting evidences
in a small number of steps.
\squishend
A key novelty is that each iteration reduces the graph in size,
and only the final iteration yields the answer and a small user-comprehensible set of explanatory evidences.
Our overarching goal is to provide end-user \textit{explain}ability to \textit{GNN} inference by \textit{iteratively} reducing the graph size, so that the final answers can indeed be claimed to be causal w.r.t. the remaining evidences, hence the name \explaignn.
A toy example of such GNN-based reduction is
% heterogeneous graph with evidences and answer candidates is shown
in Fig.~\ref{fig:graph}.
\myparagraph{Contributions}
% This work offers the following salient contributions:
We make the following salient contributions:
\squishlist
    \item Proposing a new method
    % \explaignn
    for ConvQA over heterogeneous sources, with a focus on computing explainable answers.
%    \item Presenting a flexible graph design that allows to leverage the structure among evidences from heterogeneous information sources.
    % \explaignn uses % devices a flexible heterogeneous graph, and proposes
    \item Devising a mechanism for iterative application of GNN inference
    % over a heterogeneous graph,
    % and iteratively
    to reduce such graphs until the best answers and their explanatory evidences are obtained.
%    \item Devising graph neural networks that can iteratively shrink the answering graph, to prune irrelevant information and distill the most relevant evidences and entities.
    
    \item Developing an attention mechanism which ensures that during message passing only question-relevant information is spread over the local neighborhoods.
    % \GW{what is this? something new? unless really important, could be dropped here? fewer big contributions are better than a long list} 
%
    % \item Conducting extensive experiments with the \convmix~\cite{christmann2022conversational} benchmark for ConvQA, demonstrating performance results that improve prior state-of-the-art methods by a large margin.
    % Additionally, 
    %Presenting a general user study design to investigate the explainability of ConvQA pipelines.
    % \item Conducting a crowdsourced user study on assessing user's trust in answer correctness and quality of explanatory evidences.
    % which requires a set of 
\squishend

%% file: sections/02-concepts.tex
% !TEX root = ../2023-sigir-fp-explaignn.tex
\begin{figure*} [t]
     \includegraphics[width=\textwidth]{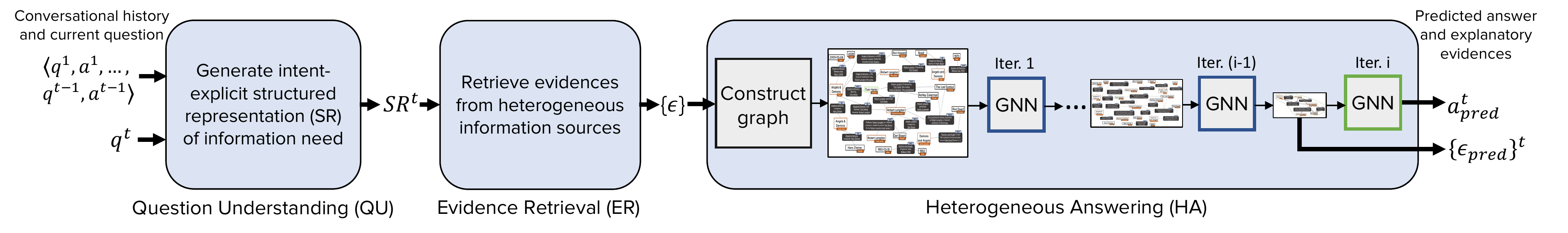}
     \vspace*{-0.8cm}
     \caption{An overview of the \explaignn pipeline, illustrating the three main stages of the approach.}
     % The output is the answer and corresponding explanation, consisting of derived intermediate results and explaining evidences.}
     \label{fig:overview}
\end{figure*}

\section{Concepts and Notation}
\label{sec:concepts}
We
% first
now introduce salient concepts and notation,
that will help understand the remainder of the paper.
Table~\ref{tab:notation} contains
the important notation.
% \vspace*{0.1cm}

%%% introduce convqa terms 
\myparagraph{Question}
A \textit{question} $q$
% can
asks about \textit{factoid} information,
like \utterance{Who wrote the book Angels and Demons?}
(intent is explicit),
% need to give one conversational example
or \utterance{How long is the novel?}
(intent is implicit).

\myparagraph{Answer}
An \textit{answer} $a$ to
% a question
$q$ can be an
entity (like \struct{Tom Hanks}), or
a 
literal (like \struct{768 pages}).

\myparagraph{Conversation}
A \textit{conversation} is a
% series
sequence
of questions and answers $\langle q^1, a^1, q^2, a^2,\dots\rangle$.
The initial question $q^1$ is complete, i.e. makes the
% whole
information need explicit.
The follow-up questions $q^t$ ($t$>$1$)
% can
may
be incomplete, 
building upon the ongoing conversational
% context,
history,
therefore leaving context information implicit.

\myparagraph{Turn}
A 
conversation
turn $t$ comprises a 
% question
$\langle q^t, a^t \rangle$ pair. %and the corresponding answer $a^t$.

\myparagraph{Knowledge base}
A curated \textit{knowledge base} is defined as a set of facts.
Each fact consists of a subject,
a
predicate,
an
object, and an optional series of $\langle$qualifier-predicate, qualifier-object$\rangle$ pairs:
% $f$ = 
\itemslist{$s$, $p$, $o$; ${qp}_1$, ${qo}_1$; ${qp}_2$, ${qo}_2$; $\dots$}.
An example fact is: \itemslist{Angels and Demons, cast member, Tom Hanks; character, Robert Langdon}.

% \myparagraph{Text corpora}
\myparagraph{Text corpus}
% please use corpus for singular, corpora plural; interchange with collections when possible
A \textit{text corpus} consists of a set of text % sentences,
documents.
% , providing information on a certain entity.

\myparagraph{Table}
A \textit{table} is a structured form of representing information,
and is typically organized
into a grid
% using
of
rows and columns.
% follow parallelism
% The individual rows often refer to an entity,
% the column header to attribute names,
% and the corresponding cell holds the attribute value.
Individual rows usually record
% record = DB term
information corresponding to specific entities,
while columns
refer to specific attributes for these entities.
% the column header to attribute names,
The row-header, column-header, and the cell hold the 
entity name, attribute name, and the attribute value, respectively.
% and the corresponding cell holds the attribute value.

\myparagraph{Infobox}
An \textit{infobox} consists of several
% infobox-entries,
entries
that are $\langle$attribute name, attribute value$\rangle$ pairs,
and provide salient information on a certain entity.
An infobox can be perceived as a special
% case
instantiation of a table,
recording 
information on a single entity,
% as a whole,
and
consisting of exactly two columns and a variable number of rows.

\myparagraph{Evidence}
An \textit{evidence} $\epsilon$ is a short text snippet expressing
% factoid
factual
information,
and can be retrieved from
% either
% dropping the either from strategic reasons
a KB, a text corpus, a table, and an infobox.
To be specific, \textit{evidences} are verbalized KB-facts, text-sentences, table-records, or infobox-entries.
% By evidence, we refer to the \textit{verbalization} of a KB fact,
% infobox-entry or table-row, i.e. a textual form of the structured artifact,
% or a sentence from text.

% A fact can be verbalized, by concatenating the individual constituents, thus constructing an evidence.
% An (attribute, value) pair can be concatenated, and prepended
% by the entity label, to form an evidence.

\myparagraph{Structured representation}
The \textit{structured representation} $SR$~\cite{christmann2022conversational}
for $q$
is an intent-explicit version of the question.
The SR represents the current question using four slots:
% of four different categories:
(i) \textit{context entity}, (ii) \textit{question entity}, (iii) \textit{relation}, (iv) \textit{expected answer type}.
% linearized
This intent-explicit representation
% is stored in
can be represented in
linear
form as a single string, using
% special separators
delimiters to separate slots
(`|' in
our
case).
% An example $SR$ for $q^3$ of the running example is:
% As an example,
The SR for $q^3$ is: %would be the following:

\begin{Snugshade}

$\langle$\,\struct{\textcolor{gray}{Angels and Demons}\,|\,\textcolor{red}{Robert Langdon}\,|\\
\indent \xspace\textcolor{blue}{who played him in the films}\,|\,\textcolor{cyan}{human}}\,$\rangle$
% \struct{[\,Angels and Demons\,|\,Robert Langdon\,|\,\\
  % who plays him in the films\,|\,human\,]}
% \noindent \struct{[Angels\&Demons|Robert Langdon|who plays him in the films|human]}
% \noindent \struct{[Angels and Demons\,|\,Dan Brown\,|\,the main character in his books\,|\,fictional character]}
\end{Snugshade}
\noindent In this example, \struct{\textcolor{gray}{Angels and Demons}} is the context entity, 
% that helps in understanding the remainder of the $SR$,
% \struct{Robert Langdon} is the question entity, - outdated
% \struct{who plays him in the films} - outdated!!
\struct{\textcolor{red}{Robert Langdon}}
% is
the question entity,
\struct{\textcolor{blue}{who played him in the films}}
% is
the relation, and \struct{\textcolor{cyan}{human}}
% is
the expected answer type.
We consider a relaxed notion of relations, in the sense of not being tied to KB terminology that canonicalizes textual relations to predicates.
This allows for softer matching in evidences,
and answering questions for which the
information cannot
easily
be represented by
such predicates.

%%% introduce notation for IGNN

%%% add a table of notations 
\begin{table} [t]
	\centering
	\caption{Notation for salient concepts in \explaignn.}
	\vspace*{-0.3cm}
	\resizebox{\columnwidth}{!}{
		\begin{tabular}{l l}
			\toprule
                \textbf{Notation}	& \textbf{Concept}\\
            \midrule
                % $C, t$ & Conversation, turn \\
                $q^t, a^t$ & Question and answer at turn $t$ \\
                $SR^t$ & Structured representation at turn $t$\\
            \midrule
                $e, \epsilon$ & Entity, evidence \\
                $E, \mathcal{E}$ & Sets of entity nodes and evidence nodes in graph \\
                $\mathcal{N}(e)$ & Evidences in $1$-hop neighborhood of entity $e$ \\
                $\mathcal{N}(\epsilon)$ & Entities in $1$-hop neighborhood of evidence $\epsilon$ \\
            \midrule
                $\boldsymbol{e}, \boldsymbol{\epsilon}$, $\boldsymbol{SR}$ & Encoding of $e$ / $\epsilon$ / $SR$ \\
                $\boldsymbol{e}^l$, $\boldsymbol{\epsilon}^l$ & Encoding of $e$ / $\epsilon$ after $l$ GNN layers\\ 
                $d$ & Encoding dimension \\
                $\alpha_{e,\epsilon}, \alpha_{\epsilon,e}$ & SR-attention of $\epsilon$ ($e$) for updating $e$ ($\epsilon$)\\
                $m_e, m_\epsilon$ & Aggregated messages passed to $e$ / $\epsilon$\\
                $s_{e}, s_{\epsilon}$ & Score for $e$ / $\epsilon$ \\
                $w_{e}, w_{\epsilon}$ & Multi-task weight for answer / evidence score prediction  \\
                $\mathcal{L}$, $\mathcal{L}_e$, $\mathcal{L}_\epsilon$ & Loss functions: total, entity relevance, evidence relevance \\
            \bottomrule
     	\end{tabular}
    }
	\label{tab:notation}
	\vspace*{-0.6cm}
\end{table}

%% file: sections/03-method.tex
\section{Overview}
\label{sec:method}
% short intro
% We devise a similar pipeline structure as 
The architecture of \explaignn (Fig.~\ref{fig:overview}) follows the pipeline of
~\convinse~\cite{christmann2022conversational}:
(i) an intent-explicit structured representation (SR) of the current information need is generated,
% constructed,
% in the question understanding (QU) phase,
(ii) evidences are retrieved from heterogeneous sources,
% in the evidence retrieval and scoring (ERS) phase,
and (iii) this large set of relevant evidences is used for 
%inferring the answer and explaining evidences for the question.
% in the final phase.
answering the question and providing explanatory evidences.
% The main contribution of this work is on the \textit{answering stage} (iii),
% but we also introduce techniques to enhance the first two stages.
% The remainder of this section 
% %will discuss the tw o initial stages in more detail.
% discusses these enhancements on (i) and (ii).
% Sec.~\ref{sec:gnn} and Sec.~\ref{sec:ignns} then present the details of the answering phase,
% the paper's main contribution.
% Fig.~\ref{fig:overview}
% gives
% % illustrates
% a
% pictorial
% overview of
% \explaignn.
% % architecture.

\subsection{Question understanding}
\label{sec:qu}
We use~\cite{christmann2022conversational} and
% The purpose of this stage is to
generate a structured representation (SR)
% , along the lines of~\cite{christmann2022conversational},
%for follow-up questions,
%that is the intent-explicit representation
%used in the remainder of the approach.
capturing the
complete
intent in the
% potentially incomplete
current question and the conversational context (the SR was loosely inspired by literature on quantity queries~\cite{ho2019qsearch}). % from the previous turns of the conversation.
%We employ the same distant supervision technique as outlined in~\cite{christmann2022conversational} to obtain training data for this stage.
For generating SRs, we leverage % a
a fine-tuned
auto-regressive
sequence-to-sequence model
(BART~\cite{lewis2020bart}). %in our experiments).
%which gets the conversational history and current question as input, and generates the SR auto-regressively.
% \GW{where is the distant supervision here? based on which training samples?}
% => \PC{not required to name it here probably, as taken from CONVINSE: we train the SR model from pure sequences of QA pairs}

\myparagraph{Preventing hallucination}
% As an enhancement,
We propose a novel mechanism to avoid
% the phenomenon of
\textit{hallucinations}~\cite{maynez2020faithfulness} in SRs.
% that language models are prone to
%for this SR generation.
%Hallucination is a common problem of sequence-to-sequence models~
% \cite{maynez2020faithfulness}:
% occasionally, outputs are generated that are completely 
% unrelated to the given input.
%\GW{better wording: completely useless as cues for answering the question !?}
For $q^3$ of the running example, 
the trained
% BART
model
could generate
\struct{\textcolor{red}{Robert de Niro}} as the (topically unrelated) question entity of the output SR
\noindent
$\langle$\,\struct{\textcolor{gray}{Dan Brown}\,|\,\textcolor{red}{Robert de Niro}\,|\,\textcolor{blue}{who played him in the films}\,|\,\textcolor{cyan}{human}}\,$\rangle$.
%\GW{generate the SR, not some answer entity? that is, the above four-tuple is generated and having de Niro in the question-entity slot is the hallucination !??? this would make sense, but then the text needs to be rephrased!?}
%This can negatively impact the rest of the pipeline, and l
%Therefore, even rare instances of hallucination can be troublesome in ConvQA systems.
This would lead the entire QA system astray.
% , and could therefore confuse end users limiting their trust in the system's predictions.

The SR is supposed to represent the information need on the \textit{surface level},
and therefore expected to use the vocabulary present in the input
(the conversational history and current question).
% Hence, it is fairly easy
% better than trivial/easy - just a flow of ideas
This makes it possible
to identify hallucinations:
output words
% (other than for the answer-type slot)
that are absent from the entire conversation so
% are,
far,
indicate such a situation.
%output words  not present in the input, it was hallucinated.
%The idea to prevent hallucinations is then as follows:
To fix this, we generate the top-$k$ SRs ($k$=$10$ in experiments),
and choose the highest-ranked SR that does not include any hallucinated words. 
Note that the expected answer type is an exception here:
it may, by design, not be present in the input.
So we remove this slot before performing the hallucination check.
Answer types are often not made explicit but substantially help the QA system~\cite{saharoy2022question}.
% Further, mapping expected answer types to a moderated KB taxonomy are useful in ambiguous cases for answer granularity.

\subsection{Evidence retrieval}
\label{sec:ers}
% structured pruning?
% Based on the SR generated in the previous stage, the goal is to retrieve relevant evidences from heterogeneous sources.
% : a curated KB, a text corpus, web tables, and Wikipedia infoboxes.

% We follow \convinse and use
% present as: how is evidence from each source obtained? more direct but of course a bit redundant
% \clocq~\cite{christmann2022beyond} is used
% for entity disambiguation and for retrieving relevant facts from the KB.
% For each entity, we also look up their Wikipedia pages
% and retrieve text sentences, 
% table rows, and infobox entries.
\textit{KB evidences} and entity disambiguations are obtained via running \clocq~\cite{christmann2022beyond} on the SR (without delimiters).
% via ,
% and retrieving facts from the KB containing these
% disambiguated
% entities.
% We use documents from Wikipedia as our text corpus, and Wikipedia
% tables and infoboxes as web tables in this work.
\textit{Text, table and infobox evidences} are obtained by mapping the disambiguated KB-entities 
% from the SR
to Wikipedia pages,
which are then parsed for extracting text-sentences, table-records, and infobox-entries corresponding to the respective entities.
Evidences from 
KB, web tables, or infoboxes,
%that naturally occur in a (semi-)stuctured manner 
% all these sources,
being natively in (semi-)structured form,
are then \textit{verbalized}~\cite{oguz2021unikqa,ma2021open}
into token sequences (as in~\cite{christmann2022conversational}). %,oguz2021unikqa,ma2021open}.
%i.e. brought into a textual form.
%The constituents of a KB fact or an infobox-entry are concatenated, using a comma as separator.
% For table rows, we include
% column headers before each individual cell, for informative context. 
%we further prepend the column header of the individual cells, to add context.
Examples can be seen inside Fig.~\ref{fig:graph}, where each evidence is tagged with its source.

% \myparagraph{Restricted ERS}
\myparagraph{Use of SR slot labels}
% Unlike~\cite{christmann2022conversational}, we make use of the SR slots for this retrieval stage.
\convinse considers 
%evidences for all entity disambiguations provided,
%and makes no difference in which slot the corresponding entity mention appears.
all entities in the SR
during retrieval,
regardless of the slot in which they appear.
In contrast,
% our method
\explaignn
\textit{restricts} the entities
%these entity disambiguations, 
by retaining only those 
%and retain only entities
%linked to
% for
mentioned within the \textit{context entity} or \textit{question entity} slots.
Evidences are then only retrieved for this restricted set of entities.
This prunes noisy disambiguations in the relation and type slots.
% In the following, we consider
%We output only evidences retrieved for this restricted set of entities,
%reducing the number of evidences retrieved, and thus the size of the initial graph in the next stage.
% and reduces the 
% size of the
% heterogeneous
% graph in the next stage.
% and also ensures better focus on the question intent.
% reducing the noise in the retrieved evidences.
% This ensures that the 

% iterative graph neural network - high-level
\section{Heterogeneous answering}
\label{sec:gnn}

We first describe heterogeneous answering graph construction (Sec.~\ref{sec:graph-constr}).
Next, we present the proposed {question-aware} GNN architecture,
consisting of the
% GNN
{encoder} (Sec.~\ref{sec:encoder}),
the {message passing} procedure (Sec.~\ref{sec:mp}),
the answer candidate {scoring} (Sec.~\ref{sec:ans-pred}),
and the {multi-task learning} mechanism for GNN training (Sec.~\ref{sec:mtl}).
% and how graph nodes are encoded
% Next we present the  GNN architecture 
% and its . 
% We then describe
% how 
% and outline a novel 
% iteratively applied to infer the final answer (Subsection 4.4) and how the GNN is trained .

% Finally, we describe
% the training and initialization process
% of these \textit{iterative GNNs} (Subsection~\ref{sec:ignn-train}).
%of ConvQA systems.
% \GW{made the respective subsections explicit - pls double-check}

\subsection{Graph construction}
\label{sec:graph-constr}
% \myparagraph{Graph construction}
Given the evidences retrieved in the previous stage,
we construct a \textit{heterogeneous answering graph}
that has two types of nodes: entities and evidences.
The graph contains
\textit{textual information} as entity labels and evidence texts, as well as the \textit{connections}
between these two kinds of nodes.
% the  \textit{textual} and \textit{structural} information
% within and between evidences.
% and corresponding entities.
%We propose the following graph structure:
% The graph has nodes for evidences and entities.
Specifically, an entity node $e$ is connected to an evidence node $\epsilon$,
if $e$ is mentioned in $\epsilon$. There are no direct edges between pairs of entities, or pairs of evidences.
An example heterogeneous graph is shown in Fig.~\ref{fig:graph}.

\myparagraph{Inducing connections between retrieved evidences} Shared entities
% are canonicalized via a KB,
% and
are the key elements that induce \textit{relationships} between the initial
% unstructured
plain \textit{set} of retrieved evidences.
%As an initial step of the graph construction,
%such entities
% have to be
%are
%identified from evidences,
%deriving a set of
So one requires % annotation of entities
entity markup
on the verbalized evidences coming from the different sources, 
that are %. These entities are
% subsequently
grounded
% linked
to a KB
for \textit{canonicalization}.
% over
% across
% all sources.
% Evidences from KB facts are
Note that during evidence verbalization, original formats are not discarded.
Thus, for KB-facts, entity mappings are already known.
% as they follow the $\langle s, p, o\rangle$ structure with optional $\langle qp, qo \rangle$ pairs (Sec.~\ref{sec:concepts}). 
% Similarly, entities in table and infobox evidences are also known from their original formats.
For text, table, and infobox evidences from Wikipedia, % href
we link anchor texts
% are linked
to their referenced
% Wikipedia
entity
pages.
These are then mapped to their corresponding KB-entities. % in the KB.
In absence of anchor texts,
% high-precision
named entity recognition and disambiguation systems can be used~\cite{li2020efficient,hoffart2011robust,van2020rel}.
Dates and years that appear in evidences are detected using regular expressions, and are added as entities to the graph as well.
% First,  $\langle$mention, entity$\rangle$ pairs
% are identified
% for each evidence, with entities
% %These entities are grounded
% linked to a KB,
% for \textit{canonicalization} across all sources.
% % These
% % sets
% % dictionary
% % are important for constructing a
% % graph
% % with canonicalized entities.
% % later on in the \explaignn pipeline,
% % since it remembers the entities mentioned in an evidence.
% For evidences from the KB, 
% one can naturally obtain these $\langle$mention, entity$\rangle$ pairs
% from the corresponding KB facts.
% For evidences obtained from Wikipedia, href anchors 
% %entity mentions to other Wikipedia pages, and map these to KB entities using dictionaries.
% are used for mapping mentions to KB-entities:
% the href anchor links referring to Wikipedia pages are mapped to the corresponding KB-entities.
% % via a dictionary.
% %Similar methods can be used for 
% % and this is used also for tables and infoboxes from Wikipedia.
% In the absence of such anchors, i.e. when using sources different
% from Wikipedia, \textit{high-precision} named entity recognition and detection (NERD)
% systems~\cite{li2020efficient,hoffart2011robust,van2020rel} could be used to
% detect 
% % KB
% entities in evidences.
% Specifically, we use \GW{??? which system/tool?}.
% rel in sigir
%could be used to detect entities in evidences.
% obtaining  pairs is
% straightforward.
In Fig.~\ref{fig:graph},
entity mentions are underlined
within evidences.
% what is the mention then? Need to say some analogous item for infobox and tables
%%%GW: the following was already said before
%Once these pairs have been identified for a given evidence,
%the evidence and the mentioned entities are added
%as graph nodes. The evidence
%is then connected
%to the corresponding entities.

% Note that this flexible graph design allows to easily
% integrate evidences from additional information sources (e.g. a News corpus)
% % can
% % be integrated easily,
% by connecting them with the corresponding entities in the graph.
% without interfering with the overall graph structure.
%Thus, it would be feasible to add new information sources to an existing graph.
% given a functionality that identifies mentioned entities in the provided evidences
% (e.g. NERD for textual sources, as discussed earlier).
%Similarly, single sources can be dropped
%by removing all evidences obtained from the specific source,
%without destroying the general graph structure.
% Note that the graph can be easily
% updated in an incremental manner,
% adding evidences and corresponding entities from additional sources,
% or dropping nodes and their incident edges.

% Since GNNs can not be efficiently applied on thousands of nodes,
% For efficiency, the initial graph is instantiated using the top-$500$ evidences as ranked using BM25 scoring.
% For applying BM25, the evidence is used as document and the SR as query.
% Having a maximum of $500$ evidences per graph
% also helps to batch graph instances during training and inference.

\subsection{Node encodings}
\label{sec:encoder}
% GNNs essentially pass messages 
GNNs
% are essentially used to
incrementally update
node encodings within local neighborhoods,
leveraging {message passing} algorithms.
However, these node encodings have to be initialized first
% which is done 
using an encoder.

\myparagraph{Evidence encodings}
For the initial encoding of the nodes, we make use of cross-encodings~\cite{reimers2019sentence}
(originally proposed for sentence pair classification tasks in~\cite{devlin2019bert}).
The evidence text, concatenated with the SR,
% and separated by a separator token,
is fed into a pre-trained language model.
% We used DistilRoberta\footnote{\url{https://huggingface.co/distilroberta-base}}, which we found to perform slightly better than DistilBERT~\cite{sanh2019distilbert} (the distillation procedure is the same).
By using SR-specific cross-encodings, we ensure that the node encodings capture the information relevant for the current question, as represented by the SR.
The encodings obtained for the individual tokens are averaged, yielding the initial evidence encoding $\boldsymbol{\epsilon^0} \in \mathbb{R}^{d}$.
% to be used within the message passing.
% \GW{there are many forward references to the GNN, but so far the text only refers to the graph and does never explicitly define the GNN ???}

\myparagraph{Entity encodings}
The entity encodings are derived analogously, using cross-encodings with the SR.
%to obtain question-relevant representations.
%However, for the entities, 
%we further 
We
% add
further
append
the
%the corresponding 
KB-type of an entity to the entity label with a separator,
% (again separated using a separator token),
before feeding the respective token sequence into the language model.
% DistilRoberta.
Including entity types is beneficial, and often crucial, for the reasons below:
% in several ways:
% \squishlist
% \item[(i)]

\noindent (i)
% When predicting the final answer, 
The cross-encoding can leverage the attention between the \textit{expected answer type} in the SR and the \textit{entity type}, which can be viewed as a soft-matching between the two. %  of the answer type;
% can be soft-matched
% against the expected answer type slot of the SR,
% to 
% % %% PC: weakened a bit, since there is no explicit "pruning"
% % % prune out
% identify
% type-incorrect answer
% candidates.
% % \item[(ii)] 

\noindent (ii) When there are multiple entities with the same label, the entity type may be a discriminating factor downstream for each of these entities (e.g., there are three entities of different types named \phrase{Robert Langdon} in Fig.~\ref{fig:graph}).
% \item[(iii)]

\noindent (iii) For long-tail entities, the entity type can add decisive informative value to the plain entity label.
% \squishend

Analogous to evidence nodes, the encodings of individual tokens are averaged to obtain the entity encoding $\boldsymbol{e^0} \in \mathbb{R}^{d}$.

\myparagraph{SR encoding}
The SR is also encoded via the language model,
% DistilRoberta
averaging the token encodings to obtain the SR encoding $\boldsymbol{SR} \in \mathbb{R}^{d}$.

Note that the same language model is used for the
initial
encodings of evidences, entities and the SR. % alike.
The parameters of this language model are updated during GNN training, to ensure that the encoder
adapts to the syntactic structure of the SR.
%within the initial layers.
% Throughout the
% In the
% remainer of this section, items in bold (e.g. \textbf{SR}) indicate their encodings.

\subsection{Message passing}
\label{sec:mp}
% \myparagraph{Message passing}
% \GW{is this MP for training? or for training and inference? pls clarify this! in the first case, I would change the subsection title into "Message passing for GNN training"!?}
%
The core part of the GNN is the \textit{message passing}~\cite{vakulenko2019message,yasunaga2021qa} procedure.
In this step, information is propagated
% across
among
neighboring nodes,
leveraging the graph structure.
Given our graph design, in each message passing step information is shared between evidences and the connected entities.
Again, we aim to focus on question-relevant information~\cite{gao2022heteroqa,brody2022attentive}, as captured by the SR,
instead of spreading general
% -purpose
information within the graph.
Therefore, we propose to weight the messages of neighboring
entities using a novel \textit{attention mechanism}, that re-weights
the messages by their question-relevance, or
equivalently,
% in this case,
their {SR-relevance}.
% (question-relevant GNN attention was also explored in the context of community QA~\cite{gao2022heteroqa}).
% by Gao et al.
This \textit{SR-attention} is computed by $\alpha_{\epsilon, e}^l$ ($\in \mathbb{R}$):

\begin{equation}
    \alpha_{\epsilon, e}^l =
    % \sigma_{\mathcal{N(\epsilon)}}
    % {\mathrm{softmax}} 
    % 
    \underset
    {\mathcal{N(\epsilon)}}
    {\mathrm{softmax}}
    \Big( \mathrm{lin}_{\alpha_{\epsilon}}^l(\boldsymbol{e}^{l-1}) \cdot \boldsymbol{SR} \Big)
    = \frac{\mathrm{lin}_{\alpha_{\epsilon}}^l(\boldsymbol{e}^{l-1}) \cdot \boldsymbol{SR}}{ \sum\limits_{e_i \in \mathcal{N(\epsilon)}} \mathrm{lin}_{\alpha_{\epsilon}}^l(\boldsymbol{e}_i^{l-1}) \cdot \boldsymbol{SR}}
    \label{eq:sr-att-evs}
\end{equation}
where we first project the entity encodings using a linear transformation ($\mathrm{lin}_{\alpha_\epsilon}^l: \mathbb{R}^d \rightarrow \mathbb{R}^d$),
and then multiply with the SR encoding to obtain a score.
The {softmax} function is then applied over all
% such scores for
entities neighboring a respective evidence ($e_i \in \mathcal{N}(\epsilon)$).
Thus, an entity can obtain different SR-attention scores
for each evidence, 
% since it depends on the scores
depending on the scores
of other neighboring entities.

The messages passed to $\epsilon$ are then aggregated, weighted by the respective SR-attention,
and projected using another linear layer:
\begin{equation}
    \boldsymbol{m}_\epsilon^l = \mathrm{lin}_{m_\epsilon}^{l} \bigg( \sum_{e \in \mathcal{N}(\epsilon)} \alpha_{\epsilon, e}^l \cdot \boldsymbol{e}^{l-1} \bigg)
\end{equation}
where $\mathrm{lin}_{m_\epsilon}^{l}$ is the linear layer ($\mathrm{lin}_{m_\epsilon}^{l}: \mathbb{R}^d \rightarrow \mathbb{R}^d$).

The updated evidence encoding is then given by
adding
the evidence encoding
from the previous layer ($\boldsymbol{\epsilon}^{l-1}$), and the messages passed from the neighbors ($\boldsymbol{m}_\epsilon^l$), activated by a ReLU function:
\begin{equation}
    \boldsymbol{\epsilon}^{l} = \mathrm{ReLU} \big( \boldsymbol{m}_\epsilon^l + \boldsymbol{\epsilon}^{l-1} \big)
\end{equation}
% Here, $\sigma$ is the activation function (we used a standard ReLU activation function in all cases).
% Intuitively,
The intuition here is that in each evidence update, the question-relevant information held by neighboring entities is passed on
to an evidence, and then incorporated in its encoding.
% After multiple GNN layers, the information spreads further in the graph.

The process for updating the entity encodings is analogous, but makes use of different linear transformation functions.
% $linear_{ev-att}$ and $linear_{ev-msg}$.
The SR-attention $\alpha_{e, \epsilon}^l$ of evidences for an entity $e$ is obtained as follows:
\begin{equation}
    \alpha_{e, \epsilon}^l =
    \underset
    {\mathcal{N}(e)}
    {\mathrm{softmax}}
    \Big( \mathrm{lin}_{\alpha_{e}}^{l} ( \boldsymbol{\epsilon}^{l-1}) \cdot \boldsymbol{SR} \Big)
\end{equation}
where $\mathrm{lin}^l_{\alpha_{e}}$ ($\mathbb{R}^d \rightarrow \mathbb{R}^d$) is the linear transformation function.
Here, the softmax function is applied over all
% evidence scores for
evidences surrounding the respective entity (i.e. $\epsilon_i \in \mathcal{N}(e)$).
Again, the messages passed to an entity $e$ are weighted by the respective SR-attention,
and projected using a linear layer ($\mathrm{lin}^l_{m_e}: \mathbb{R}^d \rightarrow \mathbb{R}^d$):
\begin{equation}
    \boldsymbol{m}_e^l = \mathrm{lin}_{m_e}^{l} \bigg( \sum_{\epsilon \in \mathcal{N}(e)} \alpha_{e, \epsilon}^l \cdot \boldsymbol{\epsilon}^{l-1} \bigg)
\end{equation}
The updated entity encoding is then given by:
\begin{equation}
    \boldsymbol{e}^{l} = \mathrm{ReLU} \big( \boldsymbol{m}_e^l + \boldsymbol{e}^{l-1} \big)
\end{equation}

These message passing steps are repeated $L$ times,
i.e. the GNN has $L$ layers.
Within these layers the question-relevant information is spread over the graph.
% Intuitively,
Basically, nodes in the graph learn about their question relevance,
based on the surrounding nodes and their relevance,
and capture this information in their node encodings.
% Note that the shown equations are crafted for explanatory purposes.
% The implementation makes use of efficient matrix multiplications,
% leveraging the adjacency matrix instead of iterating through neighborhoods,
% and processes multiple instances within a batch.

\subsection{Answer score prediction}
\label{sec:ans-pred}
% Inferring answers
Scoring answer candidates
makes use of the node encodings obtained after $L$ message passing steps.
We model the answer prediction as a {\em node classification} task~\cite{sun2018open, jia21complex, yasunaga2021qa}, by computing an answer score
for each entity node with consideration of
their question relevance as captured within the node encodings.
%making use of the question-relevant information held
%within its node encoding.
The computation of
the answer score $s_e$
is similar to %computing the 
the technique used for 
computing the
SR-attention of an entity:
\begin{equation}
    s_e = \underset
    {E}
    {\mathrm{softmax}}
    \Big( \mathrm{lin}_{e}(\boldsymbol{e}^{L}) \cdot \boldsymbol{SR} \Big)
\end{equation}
We project the entity encoding using a linear layer ($\mathrm{lin}_{e}: \mathbb{R}^d \rightarrow \mathbb{R}^d$),
and multiply the projected encoding with the encoding of the SR.
The softmax function is applied 
over
% \GW{among? do you mean: over?}
all entity nodes (i.e. $e_i \in E$).
% in the graph.
%, $E$.
%avoid unnecessary notation when it is not needed at all

% The entity that obtains the highest score,
% is our predicted answer $pa^t$.
% \begin{equation}
%     pa^t = \arg \max_{e \in E} s(e, SR) 
% \end{equation}

\subsection{Multi-task learning}
\label{sec:mtl}
% introduce different MTL strategies (only pruning, equal weights, RWL,...)
Our training data for the GNN consists of $\langle$graph, answer$\rangle$ pairs.
The gold answer is
% in our setup,
always an entity or a small set of entities.
Consequently, the positive training data is sparse:
there can be hundreds of entities in the graph
but only one gold answer.

To better use our training data, 
we propose a
% novel
\textit{multi-task learning} (MTL)~\cite{shen2019multi, kim2019gaining} approach.
Given a GNN, we pose two complementary node classification tasks:
(i) the answer prediction, and (ii) the prediction of evidence relevance.
Evidences connected to gold answers are viewed as relevant,
and others as irrelevant.
Our method learns to predict a relevance score $s_\epsilon$ for each evidence node $\epsilon$, analogous to the answer score prediction:
\begin{equation}
    s_\epsilon = 
    \underset
    {\epsilon \in \mathcal{E}}
    {\mathrm{softmax}}
    \Big( \mathrm{lin}_{\epsilon}(\boldsymbol{\epsilon}^{L}) \cdot \boldsymbol{SR} \Big)
\end{equation}
where $\mathcal{E}$ is the set of all evidence nodes (and $\mathrm{lin}_{\epsilon}: \mathbb{R}^d \rightarrow \mathbb{R}^d$).

For both tasks, answer prediction and evidence prediction, we use binary-cross-entropy over the predicted scores as loss functions:
$\mathcal{L}_e$ and $\mathcal{L}_\epsilon$, respectively. The final loss used for training the GNN is then defined as a weighted sum:
\begin{equation}
    \mathcal{L} = w_e \cdot \mathcal{L}_e + w_\epsilon \cdot \mathcal{L}_\epsilon
\end{equation}
where $w_e$ and $w_{\epsilon}$ are hyper-parameters 
%for the weights
to control the multi-task learning,
and are chosen such that $w_e + w_{\epsilon} = 1$.

% This MTL framework not only makes smarter use of the available training data, but also
% enables the the GNN models suitable to incorporate question relevance of evidences.
% \myparagraph{GNN training and inference}
% \myparagraph{GNN inference}

\begin{figure} [t]
     \includegraphics[width=0.9\columnwidth]{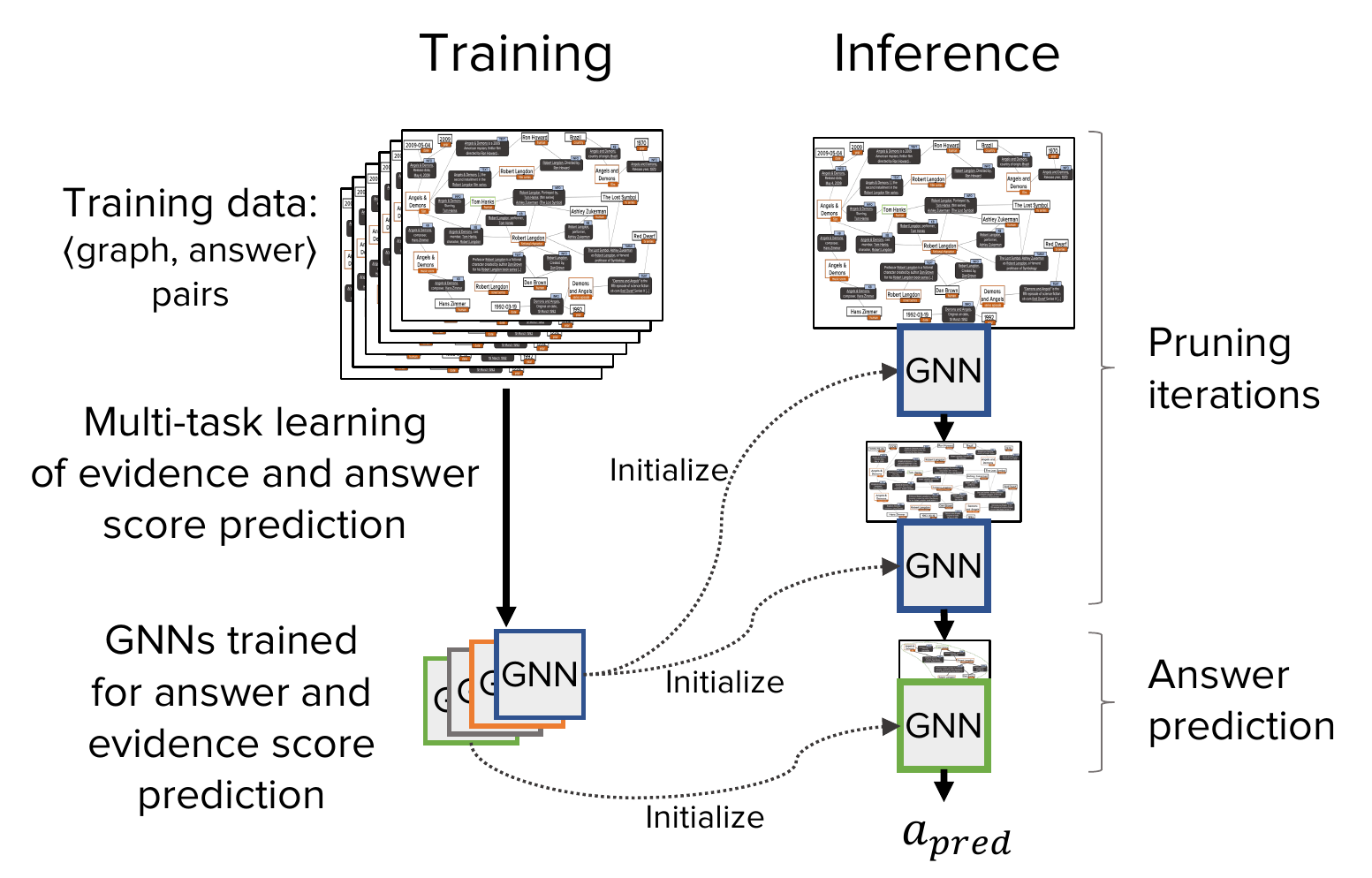}
     \vspace*{-0.4cm}
     \caption{Training of and 
     inference with
     iterative GNNs.}
     \label{fig:ignn}
     \vspace*{-0.4cm}
\end{figure}

% explain intuition behind iterative procedure
% give some examples
% \GW{again, not fully clear if this refers to training or inference or both? consider adding this to subsection title!}
%

The described GNN architecture 
can then be trained for
% so far
predicting scores of answer candidates and evidences,
and used
% So in general, the GNN can also be used directly
for inference
on the whole input graphs
% to predict the answer
in \textit{one shot}.
% , predicting the answer.
% with 
% $500$
% hundreds of evidences and entities,
% to predict the final answer
% in \textit{one shot}.
% \myparagraph{Motivation}

\section{Iterative Graph Neural Networks}
\label{sec:ignns}

We now outline 
how
% trained
we use
trained
GNNs % are used
% for 
% initializing
% an iterative mechanism 
for iteratively reducing the graph at inference time.
Specifically, we comment on the benefits that such iterative GNNs have for robustness,
explainability and efficiency. % (Sec.~\ref{sec:ignns}).

\myparagraph{Drawbacks of a one-shot prediction}
There are several drawbacks of
predicting the answer
from the full graph
\textit{during inference}:
% in such a one-shot manner:

\noindent(i) Directly predicting the answer from hundreds of answer candidates is
% a
non-trivial,
% task,
and node classification may struggle to manifest
fine-grained differences between such candidate answers in their encodings.
This can negatively impact the \textit{robustness} of the method on unseen data.
% (substantiated by results in Sec.~\ref{sec:analysis}).
% , i.e. in more challenging settings
% (we will present evidence for this claim in Sec.~\ref{sec:analysis}).

\noindent(ii) Further, if the answer is predicted from the whole input graph at inference time,
this means that all nodes in the graph contribute towards
the answer prediction.
Showing the whole graph, consisting of hundreds of nodes, to explain how the
answer was derived is not practical.
The SR-attention scores could be an indicator as to
% to identify
which nodes were
more relevant,
% found more useful,
but % while the work-wise
% of the attention mechanism was designed with this intuition in mind,
attention is not always sufficient as an explanation~\cite{jain2019attention}.
Hence, \textit{answer explainability} would be limited.

\noindent(iii) Finally, 
% even when leveraging distilled language models (like DistilRoberta in this case),
obtaining cross-encodings for hundreds of nodes (entities and evidences) can be
computationally expensive, affecting the \textit{runtime efficiency} of the system.
A large fraction of these initial graph nodes might be rather irrelevant, which can
often be identified using a more light-weight (i.e. more efficient) encoder.
% thus improving the .

\myparagraph{Iterative inference of GNNs}
To overcome these drawbacks,
we propose \textit{iterative GNNs}:
% (\textit{IGNNs}):
% GNN 
% mechanism.
instead of predicting the answer in
% a
one shot,
% manner,
we
% devise to
\textit{iteratively apply trained GNNs} of
the outlined architecture \textit{during inference}.
% we make use of the flexibility provided by the 
% heterogeneous graph design and the multi-task learning framework.
% We apply the same GNN architecture outlined above
% \textit{iteratively} on the input graph.
The key idea is to shrink the graph after each iteration.
This can be done 
% leveraging
% making use of
via the evidence scores $s_\epsilon$
predicted by the GNNs, to identify the most relevant evidences.
These evidences, and the
% corresponding
connected
entities, are used to initiate
the graph given as input to the next iteration.
In the final iteration, the answer is predicted from the reduced graph only.

Fig.~\ref{fig:ignn} illustrates training and inference with iterative GNNs.
Note that these GNNs are still trained
% for predicting answer
% and evidence relevance scores (as defined by the MTL weights),
on the full graphs in the training data, and
are
run
% on reduced graphs
iteratively
only at inference time.
This
% outlined
iterative procedure is
% applicable,
feasible
as
% because
the proposed GNN architecture
is inherently independent of the input graph size:
the same GNN trained on $500$ evidences can be applied on a graph with $100$, $20$, or $5$ evidences.
The GNNs essentially learn to spread question-relevant information within local neighborhoods,
which is 
not only
required for large graphs with hundreds of evidences, but also for smaller graphs with a handful of nodes.
Further, this iterative procedure is tractable
only with the flexibility of scoring both
entities and evidences in the graph,
using the same GNN architecture.
% Further, the MTL paradigm is a key component
% that inherently enables to score entities and evidences,
% allow the iterative inference feasible.
% enabling the shrinking of the graph.

\myparagraph{Enhancing robustness}
% With this iterative procedure, the task complexity the trained GNN model has
% to deal with at a specific iteration, is reduced during inference,
% compared 
Within each iteration,
the task complexity is decreased
compared to the task complexity the
original
GNN was trained on.
For example, the GNN was originally trained for predicting small answer sets
and relevant evidences from
% thousands
several hundreds of nodes,
but 
in the iterative setup,
rather
needs
to identify 
the top-$100$ evidences during inference.
Thus, the 
GNN model now
has to be \textit{less discriminatory} at inference time
than during training.
% Different from the one-shot procedure,
% the answering task is split into multiple sub-tasks.
This can help improve \textit{robustness}.
% of the method.
% when applied on unseen data (e.g. other datasets).

\myparagraph{Facilitating explainability}
% Another
A primary
benefit of the iterative mechanism is that the intermediate graphs
can be used to better understand how answer prediction works via a GNN.
% , and therefore enhance
% the explainability.
Showing all the information
% represented
contained
in the 
original
% full
input graph with hundreds of 
% nodes
entities and evidences
to the user is not practical:
we can iteratively derive a small set of evidences (say five),
from which the answer is predicted.
These
evidences
% provided
can then be presented 
to the end user,
enhancing
% the\
\textit{user explainability}.

\myparagraph{Improving efficiency}
To facilitate the runtime \textit{efficiency} of the answering process,
we refrain from encoding entities via cross-encodings in the \textit{shrinking} (or \textit{pruning}) iterations.
Instead, we initiate the entity encodings, using
a sum of the surrounding evidences, weighted by their question-relevance (i.e. their SR-attention):
\begin{equation}
    \boldsymbol{e}^0 = \sum_{\epsilon \in \mathcal{N}(e)} \alpha_{e,\epsilon} \cdot \boldsymbol{\epsilon}^{0}
    \label{eq:alt-encodings}
\end{equation}
where the SR-attention $\alpha_{e,\epsilon}$
% of an evidence $\epsilon$
is computed as in Eq.~\ref{eq:sr-att-evs}, employing a different linear projection.
This can be perceived as obtaining \textit{alternating encodings} of entities:
the initial evidence encodings are used
% for the initial
to initialize
entity encodings (inspired by~\cite{chatterjee2022bert}).
% Note the similarity to work in entity ranking that encodes entities using related Wikipedia sentences~\cite{chatterjee2022bert}.
The message passing would then proceed as outlined in Sec.~\ref{sec:mp}.
% Note that the GNN has to be trained for incorporating these
% \textit{alternating encodings} for entities.
% We found this \textit{alternating encoding} of entity nodes sufficient
% in the initial iterations.
% to distill the important parts of the graph, while it can substantially
% improve the runtime efficiency.
% Only in the final answering iteration,
% entities are encoded using cross-encodings.
% the full encoder is applied for the entities, making use of the cross-encoding.

%% dropped for space reasons + forecasting experiments
% The alternating entity encoder can even stengthen the usage of the graph structure
% in the initial iterations, since the GNN can not solely rely on the node content,
% which is a desirable property.

\myparagraph{Instantiation}
% \subsection{Instantiation}
% \label{sec:ignn-train}
% which instances pruned etc (instances without an answer)
% only-
% The most straightforward way to train the iterative GNNs would be to train one dedicated GNN for each iteration,
% that can be fine-tuned for the specific graph characteristics of the individual iteration (such as graph size or connectivity).
% However, this would result in a lot of model parameters, and also hyperparameter choices within the iterations.
% 
% We propose a different training routine: 
We train several one-shot GNNs of the architecture outlined above, using different
% multi-task learning strategies.
% These vary the
weights
% put
on the answer prediction and evidence relevance prediction tasks ($w_e$ and $w_\epsilon$ respectively) in the MTL setup.
Further, we train GNNs using either cross-encodings or alternating encodings for entities.
Training is conducted on the full input graphs present in the training set.
We then simply instantiate all \textit{pruning iterations} with the GNN that obtains the
best evidence prediction performance on the graphs in the development (dev) set.
Similarly, we use the trained GNN that obtained the best answering performance
on the dev set to initiate the final \textit{answering iteration}.
Finally,
the answer
% and the explanatory evidences 
predicted by the system
% , $a_{pred}$,
is given by:
\begin{equation}
    a_{pred} = 
    \arg \max_{e \in E} s_e
\end{equation}
This is shown to the end user, together with the explanatory evidences $\{\epsilon_{pred}\}$ (see outputs in Fig.~\ref{fig:overview}), that are simply the set of evidences used in the answer prediction step.

%% file: sections/04-experiments.tex
\section{Results and insights}
\label{sec:experiments}

\subsection{Experimental setup}
\label{sec:setup}

\myparagraph{Dataset}
We train and evaluate \explaignn on the \convmix~\cite{christmann2022conversational}
benchmark, which was designed for ConvQA over heterogeneous information sources.
The dataset has $16{,}000$ questions (train: $8{,}400$ questions, dev: $2{,}800$ questions, test: $4{,}800$ questions),
within $3{,}000$ conversations of five ($2{,}800$) or ten turns ($200$, only used for testing).

% \subsection{Metrics}
% \vspace*{0.2cm}
\myparagraph{Metrics}
For accessing the answer performance, we use \textit{precision at 1} (\textbf{P@1}),
as suggested for the \convmix benchmark.
To investigate the ranking capabilities of different methods in more detail,
we also measure the
\textit{mean reciprocal rank} (\textbf{MRR}),
and \textit{hit at 5} (\textbf{Hit@5}).
The \textit{answer presence} \textbf{(Ans. pres.)} is the fraction of questions for which a gold answer is present in a given set of evidences.
% metrics.
% Throughout the experiments, we may also measure the \textbf{runtime} of the answering stage,
% or the \textbf{answer presence} within a set of evidences.

% \subsection{Baselines}
% \vspace*{0.2cm}
\myparagraph{Baselines}
We compare \explaignn with the state-of-the-art method on the \convmix dataset,
\convinse~\cite{christmann2022conversational}.
\convinse leverages a Fusion-in-Decoder (FiD)~\cite{izacard2021leveraging} model for obtaining
the top answer, which is designed to generate a single answer string.
In~\cite{christmann2022conversational}, the ranked entity answers are then derived
by collecting the top-$k$ answer candidates with the
highest surface-form match with respect to Levenshtein distance with
the generated answer string.
This procedure is somewhat limiting when measuring metrics beyond the first rank (i.e. MRR or Hit@5).
Therefore, we enhanced the FiD model
to directly generate top-$k$ answer strings, and then consider the answer candidate with
the highest surface-form match for each such generated string (\textbf{top-$k$ FiD}),
for fair comparison.
We further compare with baselines proposed in~\cite{christmann2022conversational} using question completion and resolution,
and use the values reported in~\cite{christmann2022conversational} for consistency.
Note that these methods transform a conversational question to a self-sufficient form, and still need to be coupled with retrieval and answer generation modules.
% the \convinse paper:
% for answering.

% \subsection{Configurations}
% \vspace*{0.2cm}
\myparagraph{Configurations}
% We make use of the code provided in the \convinse repo\footnote{\url{https://github.com/PhilippChr/CONVINSE}},
% to initiate the Wikipedia retriever and KB access.
The QU and ER stages were initialized using the \convinse~\cite{christmann2022conversational} code and data:
we used Wikidata as the KB, and made use of the same version (31 January 2022) as earlier work.
% \footnote{The dump is hosted on \url{https://clocq.mpi-inf.mpg.de/documentation}.}.
The Wikipedia evidences were taken from here\footnote{\url{http://qa.mpi-inf.mpg.de/convinse/convmix_data/wikipedia.zip}}, whenever applicable, and retrieved on-the-fly 
% and cached
otherwise.
This ensures that results are comparable with the results provided in earlier work.
\convmix has $\simeq 3\%$ of yes/no questions: these are out of scope for \explaignn. To be able to report numbers on the full benchmark, we follow previous work~\cite{christmann2022conversational} and detect such questions as starting with an auxiliary verb, and answer \phrase{yes} to these.

The GNNs were implemented from scratch via PyTorch. %\footnote{\url{https://pytorch.org}}.
% using the DistilRoberta model provided by Hugging Face\footnote{\url{https://huggingface.co/distilroberta-base}} to initialize the encoder. 
We used DistilRoBERTa provided by Hugging Face\footnote{\url{https://huggingface.co/distilroberta-base}} as encoder, which we found to perform slightly better than DistilBERT~\cite{sanh2019distilbert} (the distillation procedure is the same).
DistilRoBERTa has an embedding dimension of $d$=$768$.
The GNNs were trained for $5$ epochs.
We chose an epoch-wise evaluation strategy, and kept the model that achieved the best performance on the dev set.
We found three layer GNNs ($L$=$3$) most effective.
With our graph schema of alternating entities and evidences, using an odd number of layers allows for information from evidences to reach relevant entities in their immediate (one hop) and slightly distant neighborhoods (three or five hops, say).
% on the dev set.
% Intuitively, this means that information aggregated in the entity encoding of the final layer can come from the nodes that are $3$ hops away.
AdamW was used as optimizer, using a learning rate of $10^{-5}$, batch size of $1$, and a weight decay of $0.01$.

We also used the dev set for choosing the number of
GNN 
iterations,
and MTL weights for pruning and answering.
The number of iterations was set to $i$=$3$.
The GNN that maintains the highest answer presence among the top-$5$
% predicted
evidences was chosen for instantiating the pruning iterations (alternating encodings for entities, $w_{e}$=0.3, $w_{\epsilon}$=0.7),
and the GNN obtaining the highest P@1 for the
% final
answer prediction (cross-encodings for entities, $w_{e}$=0.5, $w_{\epsilon}$=0.5).
In case there are more than $500$ evidences, we retain only the top-$500$ obtained via BM25 scoring as input to the answering stage.
% Models trained for pruning purposes
% were chosen by their answer presence among the top-$5$ predicted evidences,
% the models trained for answering purposes were chosen by their P@1 metric.
% We chose $i$=$3$, pruning from $500$ to $100$ to $20$ evidences
% via a GNN based on an alternating encoding ($w_e$=$0.3$, $w_\epsilon$=$0.7$),
% and then predicting the answer with a GNN trained using $w_e$=$0.5$ and $w_\epsilon$=$0.5$.
A single GPU (NVIDIA Quadro RTX 8000, 48 GB GDDR6) was used to 
train and evaluate the models.

%% file: sections/05-results.tex
% !TEX root = ../2023-sigir-fp-explaignn.tex
% \subsection{Results and insights}
% \label{sec:res}
% This section will first present the main experimental results
% on the \convmix test set, and then analyze the answering performance
% % of the pipeline
% of different variations
% and settings in more depth.

\subsection{Key findings}
\label{sec:main-res}

This section will present the main experimental results
on the \convmix test set.
All provided metrics are averaged over all \textit{questions} in the dataset.
The best method for each column is shown in \textbf{bold}.
Statistical significance over the best baseline is indicated by an asterisk (*),
and is measured via paired $t$-tests for MRR,
and McNemar's test for binary variables (P@1 or Hit@5), with $p$
% -value
$< 0.05$ in both cases.

\begin{table} [t]
    \caption{Comparison of answering performance on the \convmix~\cite{christmann2022conversational} test set, using \textit{gold answers} $\{a_{gold}\}$ in the history.}
    \vspace*{-0.4cm}
    \newcolumntype{G}{>{\columncolor [gray] {0.90}}c}
    \resizebox*{\columnwidth}{!}{
    	\begin{tabular}{l G G G}
    		\toprule
        		\textbf{Method $\downarrow$} & \textbf{P@1}  & \textbf{MRR}  & \textbf{Hit@5} \\ 
        	\midrule
        		\textbf{Q. Resolution~\cite{voskarides2020query} + BM25 + FiD~\cite{izacard2021leveraging}}
        		& $0.282$ &   $0.289$ & $0.297$  \\
          
        		\textbf{Q. Rewriting~\cite{raposo2022question} + BM25 + FiD~\cite{izacard2021leveraging}}
        		& $0.271$ &   $0.278$ & $0.285$  \\
        	% \midrule
        		\textbf{\convinse~\cite{christmann2022conversational} (original)}
                & $0.342$ &   $0.365$ & $0.386$  \\

                \textbf{\convinse~\cite{christmann2022conversational} (top-$k$ FiD)} 
        	    & $0.343$ &   $0.378$ & $0.431$  \\
            \midrule
        	    % \textbf{$\explaignn_{500}$} 
        	    % & $0.406$ &   $0.469$ & $0.550$  \\
                \textbf{$\explaignn$ (proposed)} 
        	    & $\boldsymbol{0.406}$* &   $\boldsymbol{0.471}$* & $\boldsymbol{0.561}$*  \\
                % \textbf{$\explaignn_{5}$} 
        	    % & $0.-$ &   $0.-$ & $0.-$  \\
    		\bottomrule
    	\end{tabular} 
    }
    \label{tab:main-res}
    \vspace*{-0.3cm}
\end{table}

\myparagraph{\explaignn improves the answering performance}
The main results in Table~\ref{tab:main-res}
demonstrate the performance benefits of \explaignn over
the baselines.
\explaignn significantly outperforms the best baseline
on all metrics,
illustrating the success of using \textit{iterative graph neural networks} in ConvQA. % task. % of ConvQA. % conversational question answering.
% % establishing the new state-of-the-art.
% performance.
As is clear from the method descriptions in Table~\ref{tab:main-res}, all baselines crucially rely on the generative reader
model 
of
FiD.
FiD can ingest multiple evidences to produce the answer, but fails to capture their relationships explicitly, that is unique to our graph-based pipeline.
Our adaptation of using
top-$k$ FiD instead of the default top-1 
improved the ranking capabilities (MRR, Hit@5)
of the strongest baseline \convinse. 
However, \explaignn 
still substantially improved over \convinse with top-$k$ FiD. % on both metrics.
% can still improve the Hit@5 
% by 13 total points.
% While using top-$k$ FiD results substantially
% improves the ranking capabilities (MRR, Hit@5)
% of \convinse, \explaignn 
% can still improve the Hit@5 
% by 13 total points.

\begin{table} [t] 
    \caption{Comparison of answering performance on the \convmix test set, using \textit{predicted answers} $\{a_{pred}\}$ in the history.}
    % \caption{Answer performance of ConvQA systems when using the \emph{predicted answers} for the conversation history.}
    \vspace*{-0.4cm}
    \newcolumntype{G}{>{\columncolor [gray] {0.90}}c}
    \resizebox*{\columnwidth}{!}{
    	\begin{tabular}{l G G G}
    		\toprule
    		    % & \multicolumn{3}{c}{\textbf{\convmix~\cite{christmann2022conversational}}} \\
    		     \textbf{Method $\downarrow$}	 & \textbf{P@1}  & \textbf{MRR}  & \textbf{Hit@5}\\
    		    \midrule
            	    \textbf{Q. Resolution~\cite{voskarides2020query}+ BM25 + FiD~\cite{izacard2021leveraging}}   &	$0.243$ &	$0.250$ &	$0.257$ \\
            	    \textbf{Q. Rewriting~\cite{raposo2022question}+ BM25 + FiD~\cite{izacard2021leveraging}}    &	$0.221$ &	$0.227$ &	$0.235$ \\  
        	        \textbf{\convinse~\cite{christmann2022conversational} (original)}          &	${0.278}$ &	${0.286}$ &	${0.294}$ \\
                    \textbf{\convinse~\cite{christmann2022conversational} (top-$k$ FiD)}          &	$0.279$ &	$0.308$ &	$0.351$ \\
                \midrule
                    % \textbf{$\explaignn_{500}$}        &	$0.341$ &	$0.397$ &	$0.469$ \\
                    \textbf{$\explaignn$ (proposed)}         &	$\boldsymbol{0.339}$* &	$\boldsymbol{0.398}$* &	$\boldsymbol{0.477}$* \\
                    % \textbf{$\explaignn$}          &	$-$ &	$-$ &	$-$ \\
    		\bottomrule
    	\end{tabular}
     }
    \label{tab:pred-answer}
    \vspace*{-0.4cm}
\end{table}

% \myparagraph{\explaignn can overcome failures}
\myparagraph{\explaignn is robust to wrong predictions in earlier turns}
Unlike many existing works, we also evaluated the methods in a \textit{realistic scenario},
in which the predicted answers $a_{pred}$ are used as (noisy) input
for the conversational history
instead of the standard yet impractical choice
% practice
of inserting gold answers from the benchmark.
Results are shown in Table~\ref{tab:pred-answer} (cf. Table~\ref{tab:main-res} results, that show an evaluation with gold answers).
While the performance of all methods drops
in this more challenging setting,
the trends are very similar, indicating that
\explaignn can successfully overcome failures
(i.e. incorrect answer predictions)
in earlier turns.
\explaignn again outperforms all baselines
significantly, 
including a P@1 jump from $0.279$ for the strongest baseline to $0.339$.
% with benefits for the P@1 metric of $6.2$ total points.

\begin{table} [t] 
    \caption{Effect of varying source combinations at inference time (test set). \explaignn is still \textit{trained} on all sources.}
    % Note that the \textit{same trained pipeline} is used across these experiments (\explaignn trained on all sources).}
    \vspace{-0.4cm}
    \newcolumntype{G}{>{\columncolor [gray] {0.90}}c}
    \newcolumntype{H}{>{\setbox0=\hbox\bgroup}c<{\egroup}@{}}
    	\begin{tabular}{l G G G c} 
        \toprule
            \textbf{Method $\downarrow$} & \textbf{P@1}  & \textbf{MRR}  & \textbf{Hit@5} & \textbf{Ans. pres.} \\
            \midrule
                \textbf{KB}             &  $0.363$  &  $0.427$  &  $0.511$ &  $0.617$ \\
                \textbf{Text}           &  $0.233$  &  $0.300$  &  $0.380$ &  $0.530$ \\
                \textbf{Tables}         &  $0.064$  &  $0.084$  &  $0.108$ &  $0.155$ \\
                \textbf{Infoboxes}           &  $0.256$  &  $0.302$  &  $0.362$ &  $0.409$ \\
            \midrule
                \textbf{KB+Text}        &  $0.399$  &  $0.464$  &  $0.549$ &  $0.672$ \\
                \textbf{KB+Tables}      &  $0.363$  &  $0.429$  &  $0.515$ &  $0.629$ \\
                \textbf{KB+Infoboxes}        &  $0.376$  &  $0.443$  &  $0.532$ &  $0.640$ \\
                \textbf{Text+Tables}    &  $0.235$  &  $0.305$  &  $0.392$ &  $0.540$ \\
                \textbf{Text+Infoboxes}      &  $0.309$  &  $0.369$  &  $0.445$ &  $0.572$ \\
                \textbf{Tables+Infoboxes}    &  $0.263$  &  $0.312$  &  $0.374$ &  $0.453$ \\
            \midrule
                \textbf{All sources}    &  $\boldsymbol{0.406}$  &  $\boldsymbol{0.471}$  &  $\boldsymbol{0.561}$ &  $\boldsymbol{0.683}$ \\
            \bottomrule
    \end{tabular}
    \label{tab:sources}
    \vspace*{-0.2cm}
\end{table}

\myparagraph{Heterogeneous sources improve performance}
% Another main desideratum of the \explaignn pipeline is that
% the potential of using
% heterogeneous sources can
% % successfully be incorporated.
% be successfully leveraged.
% Therefore,
% We analyze different
% % choices of
% source combinations,
% and report
% % the
% results in Table~\ref{tab:sources}.
Analysis of source combinations is in Table~\ref{tab:sources}.
The first takeaway is that the answering performance was the best
when the full spectrum of sources was used.
\explaignn can make use of the enhanced answer presence in this
case, and answer more questions correctly.
Next, the results indicate that adding an information
source is always beneficial:
the performance of combinations of two sources is in all cases
better than for the two sources individually.

\subsection{In-depth analysis}
\label{sec:analysis}

\begin{table} [t] 
    \caption{Effect of varying the multi-task learning weights \textit{when training} the one-shot GNN modules (on the dev set).}
    \vspace*{-0.4cm}
    \newcolumntype{G}{>{\columncolor [gray] {0.90}}c}
    \newcolumntype{H}{>{\setbox0=\hbox\bgroup}c<{\egroup}@{}}
    \resizebox*{\columnwidth}{!}{
    	\begin{tabular}{l G G G c G} 
        % \begin{tabular}{l G G G G G c c c} 
    		\toprule
                	 & \textbf{P@1}  & \textbf{MRR}  & \textbf{Hit@5} & \textbf{Ans. pres.} & \textbf{HA runtime} \\
    		    \midrule
                    \multicolumn{6}{c}{\textbf{$\explaignn$} ($i$=$1$: 500$\rightarrow$$a_{pred}$; \textbf{cross-encodings} for entities)}\\
                \midrule
            	    $w_e$=$1.0$, $w_{\epsilon}$=$0.0$   & $0.440$  & $0.501$  & $0.578$ & $0.229$  & $1{,}018$\,ms \\
            	    $w_e$=$0.7$, $w_{\epsilon}$=$0.3$   & $0.439$  & $0.499$  & $0.573$ & $0.573$  & $1{,}029$\,ms \\
        	        $w_e$=$0.5$, $w_{\epsilon}$=$0.5$   & $0.442$  & $0.502$  & $0.581$ & $0.583$  & $1{,}017$\,ms \\
                    $w_e$=$0.3$, $w_{\epsilon}$=$0.7$   & $0.431$  & $0.495$  & $0.572$ & $0.586$  & $1{,}013$\,ms \\
                    $w_e$=$0.0$, $w_{\epsilon}$=$1.0$   & $0.033$  & $0.041$  & $0.044$ & $0.579$  & $1{,}008$\,ms \\
                \midrule
                    \multicolumn{6}{c}{\textbf{$\explaignn$} ($i$=$1$: 500$\rightarrow$$a_{pred}$; \textbf{alternating encodings} for entities)}\\
                \midrule
                    $w_e$=$1.0$, $w_{\epsilon}$=$0.0$  & $0.417$  & $0.485$  & $0.573$  & $0.508$ & $443$\,ms \\
            	    $w_e$=$0.7$, $w_{\epsilon}$=$0.3$  & $0.410$  & $0.470$  & $0.545$  & $0.568$ & $444$\,ms \\
        	        $w_e$=$0.5$, $w_{\epsilon}$=$0.5$  & $0.404$  & $0.472$  & $0.555$  & $0.569$ & $447$\,ms \\
                    $w_e$=$0.3$, $w_{\epsilon}$=$0.7$  & $0.405$  & $0.472$  & $0.552$  & $0.589$ & $442$\,ms \\
                    $w_e$=$0.0$, $w_{\epsilon}$=$1.0$  & $0.117$  & $0.169$  & $0.221$  & $0.581$ & $449$\,ms \\
    		\bottomrule
    	\end{tabular}
     }
    \label{tab:mtl}
    \vspace*{-0.3cm}
\end{table}

\begin{table*} [t] 
    \caption{Varying the no. of iterations, pruning factors, and evidences the answer is predicted from \textit{during inference} (dev set).}
    % of the \convmix~\cite{christmann2022conversational} dataset.}
    \vspace*{-0.4cm}
    \newcolumntype{G}{>{\columncolor [gray] {0.90}}c}
    \newcolumntype{H}{>{\setbox0=\hbox\bgroup}c<{\egroup}@{}}
    \resizebox*{\textwidth}{!}{
        \begin{tabular}{l G G G c c c G}
            \toprule
                % & \multicolumn{3}{G}{\textbf{Dev set - Answering}} & \multicolumn{3}{c}{\textbf{Dev set - Answer presence}} & \textbf{Runtime} \\
                \textbf{Method $\downarrow$}	 & \textbf{P@1}  & \textbf{MRR}  & \textbf{Hit@5} & \multicolumn{3}{c}{\textbf{Ans. pres. and no. of evidences after pruning iteration}} & \textbf{HA runtime} \\ 
            \midrule
                \textbf{$\explaignn$} ($i$=$1$: 500$\rightarrow$$a_{pred}$)         & $\boldsymbol{0.442}$  &	$0.502$   &	$0.581$    & $\hspace*{1cm}-\hspace*{1cm}$  &	$\hspace*{1cm}-\hspace*{1cm}$   &	$\hspace*{1cm}-\hspace*{1cm}$ &	$1{,}017$\,ms \\
            \midrule
                \textbf{\explaignn} ($i$=$2$: 500$\rightarrow$100$\rightarrow$$a_{pred}$)  & $0.441$  &	$0.504$   &	$0.587$    & $-$  &	$-$   &	$0.687$\, | $100$ &	$744$\,ms \\
                \textbf{\explaignn} ($i$=$2$: 500$\rightarrow$50$\rightarrow$$a_{pred}$)         & $0.440$  &	$0.504$   &	$0.588$    & $-$  &	$-$   &	$0.675$\, | $50$ &	$591$\,ms \\
                \textbf{\explaignn} ($i$=$2$: 500$\rightarrow$20$\rightarrow$$a_{pred}$)         & $0.438$  &	$0.504$   &	$\boldsymbol{0.591}$    & $-$  &	$-$   &	$0.655$\, | $20$ &	$515$\,ms \\
                \textbf{$\explaignn$} ($i$=$2$: 500$\rightarrow$5$\rightarrow$$a_{pred}$)          & $0.422$  &	$0.480$   &	$0.560$    & $-$  &	$-$   &	$0.589$\, | $5$ &	$459$\,ms \\
            \midrule
                \textbf{\explaignn} ($i$=$3$: 500$\rightarrow$200$\rightarrow$100$\rightarrow$$a_{pred}$)   & $0.441$  &	$0.504$   &	$0.585$    & $-$  &	$0.694$\, | $200$   &	$0.685$\, | $100$ &	$995$\,ms \\
                \textbf{\explaignn} ($i$=$3$: 500$\rightarrow$150$\rightarrow$50$\rightarrow$$a_{pred}$)    & $0.441$  &	$\boldsymbol{0.505}$   &	$0.586$    & $-$  &	$0.693$\, | $150$   &	$0.678$\, | $50$ &	$741$\,ms \\
                \textbf{$\explaignn$} ($i$=$3$: 500$\rightarrow$100$\rightarrow$20$\rightarrow$$a_{pred}$; \textbf{proposed})  & $\boldsymbol{0.442}$  &	$\boldsymbol{0.505}$   &	$0.589$    & $-$  &	$0.687$\, | $100$   &	$0.654$\, | $20$ &	$601$\,ms \\
                \textbf{\explaignn} ($i$=$3$: 500$\rightarrow$50$\rightarrow$5$\rightarrow$$a_{pred}$)      & $0.419$  &	$0.475$   &	$0.556$    & $-$  &	$0.675$\, | $50$   &	$0.579$\, | $5$ &	$511$\,ms \\
            \midrule
                \textbf{\explaignn} ($i$=$4$: 500$\rightarrow$300$\rightarrow$150$\rightarrow$100$\rightarrow$$a_{pred}$)        & $0.441$  &	$0.504$   &	$0.587$    & $0.696$\, | $300$  &	$0.691$\, | $150$   &	$0.686$\, | $100$ &	$1{,}232$\,ms \\
                \textbf{\explaignn} ($i$=$4$: 500$\rightarrow$200$\rightarrow$100$\rightarrow$50$\rightarrow$$a_{pred}$)         & $0.440$  &	$0.504$   &	$0.585$    & $0.694$\, | $200$  &	$0.685$\, | $100$   &	$0.677$\, | $50$ &	$945$\,ms \\
                \textbf{\explaignn} ($i$=$4$: 500$\rightarrow$200$\rightarrow$50$\rightarrow$20$\rightarrow$$a_{pred}$)          & $0.436$  &	$0.500$   &	$0.584$    & $0.694$\, | $200$  &	$0.677$\, | $50 $   &	$0.652$\, | $20$ &	$769$\,ms \\
                \textbf{\explaignn} ($i$=$4$: 500$\rightarrow$100$\rightarrow$20$\rightarrow$5$\rightarrow$$a_{pred}$)           & $0.422$  &	$0.476$   &	$0.553$    & $0.687$\, | $100$  &	$0.654$\, | $20 $   &	$0.575$\, | $5$ &	$577$\,ms \\
            \bottomrule
        \end{tabular} 
    }
    \label{tab:iterations}
    \vspace*{-0.3cm}
\end{table*}

\myparagraph{Multi-task learning enables flexibility}
A systematic analysis of the effect of different multi-task learning
weights is shown in Table~\ref{tab:mtl}.
This analysis is conducted on the dev set,
% as outlined 
% and conducted for a one-shot GNN with either the default encodings,
for choosing the best GNN for pruning and answering, respectively.
% or alternating encodings for entities
Entities are either encoded via cross-encodings,
or alternating encodings (see Eq.~\ref{eq:alt-encodings}).
The results indicate the runtime benefits of using alternating encodings for entities.
% which can substantially decrease the runtime.
Further, when optimized for evidence relevance prediction ($w_{\epsilon} > 0.5$) % or $w_{\epsilon}$=1.0),
it can maintain a high answer presence (measured within top-$5$ evidences), 
% which gives evidence to the claim that more
indicating that light-weight encoders are indeed sufficient for the pruning iterations.
% to distill out the important information.
Further, we found putting equal weights on the answer and evidence relevance prediction
% tasks
to be
beneficial for answering.
% performance.
% (compared to putting the whole
% weight on the answer prediction).

% \myparagraph{\explaignn is robust to different choices for the iterations}
% \myparagraph{\explaignn is configurable for various objectives}
\myparagraph{Iterative GNNs do not compromise runtimes}
% robust to different choices for the iterations}
Table~\ref{tab:iterations} reports results of varying the 
number of iterations $i \in \{1, 2, 3, 4\}$, and the graph size in the number of evidences the answer is predicted from ($|\mathcal{E}| \in \{500, 100, 50, 20, 5\}$).
For each row, the 
% pruning factor
reduction in graph size in terms of the number of evidences considered
was kept roughly constant % at $20$-$25\%$
for consistency.
% The pruning and answering GNNs were chosen as 
% Results are shown in Table~\ref{tab:iterations}.
% First of all,
By our smart use of alternating encodings of entities in the pruning iterations,
% the number of iterations does not influence the \textit{runtime} much.
\textit{runtimes remain immune} to the number of pruning iterations (times for $i$=$4$ are not necessarily higher than those for $i$=$3$, and so on). % are kept mostly 
Rather, the runtime is primarily influenced by the size of the graph (in terms of the number of evidences) given to the final answer prediction step (compare runtimes within each iteration group).
Recall that this graph size for answer prediction can impact \textit{explainability} to end users,
if it is not small enough.
% of end users.
Notably, % \explaignn achieves 
% We found that the pipeline
% is quite robust to different choices,
% and
% achieves
\textit{performance} remains
reasonably stable in most cases.
% A typical reader could have a belief upfront that using multiple GNN iterations before an answer is predicted would drastically degrade runtimes.
% It may be expected that having multiple iterations be
% runtime would 
A key takeaway from these results is that the trained GNN models generalize well to graphs of different sizes. Concretely, while all of these models are trained on graphs established from 500 evidences, they can be applied to score nodes in graphs of variable sizes.

\myparagraph{\explaignn can be applied zero-shot}
For testing the generalizability of \explaignn,
we applied the pipeline trained on the \convmix dataset
out-of-the-box, without any training or fine-tuning,
on the \convquestions~\cite{christmann2019look} dataset.
\convquestions is a competitive benchmark for
% evaluating
ConvQA methods
operating over KBs. We test the same \explaignn pipeline in two different
modes: (i) using only facts from the KB, and (ii) using
% the full set of
evidences from all information sources.
Table~\ref{tab:convquestions} shows the results ($^\dag$ and $^\ddag$ indicate statistical significance over the leaderboard toppers \textsc{Krr}~\cite{ke2022knowledge} and \textsc{Praline}~\cite{kacupaj2022contrastive}, respectively).
In the KB-only setting, \explaignn obtains state-of-the-art
performance, reaching the highest MRR score.
Also, we found that integrating heterogeneous sources can 
improve the answer performance substantially for \convquestions, even though this benchmark was created with a KB in mind. 
% a source-specific setting.

\myparagraph{Iterative GNNs improve robustness}
In Sec.~\ref{sec:gnn}, we argued that iterative GNNs
enhance the
% generalizability,
pipeline's
robustness
over a single
GNN applied on the full graph in one shot.
While the performance of the one-shot GNN on the \convmix dev set
is comparable
(Table~\ref{tab:iterations}), we found that it cannot generalize as well to
a different dataset. When applied on \convquestions,
the respective performance of the one-shot GNN is significantly lower than for \explaignn,
in both KB-only (P@1: $0.330$ versus $0.281$) and heterogeneous (P@1: $0.363$ versus $0.318$) settings.

\begin{table} [t] 
    \caption{Out-of-the-box \explaignn, without further training or fine-tuning, on the \convquestions~\cite{christmann2019look} benchmark.}
    \vspace*{-0.4cm}
    \newcolumntype{G}{>{\columncolor [gray] {0.90}}c}
    \resizebox{\columnwidth}{!}{
    \begin{tabular}{l G G G}
        \toprule
            % & \multicolumn{3}{G}{\textbf{\convquestions~\cite{christmann2019look}}} \\
            \textbf{Method $\downarrow$}	 & \textbf{P@1}  & \textbf{MRR}  & \textbf{Hit@5}\\
        \midrule
            \textbf{\textsc{Convex}~\cite{christmann2019look}} 
            & $0.184$ & $0.200$  &	$0.219$ \\
            \textbf{\textsc{Focal Entity Model}~\cite{lan2021modeling}} 
            & $0.248$ & $0.248$  &	$0.248$ \\
            \textbf{\textsc{Oat}~\cite{marion2021structured}} 
            & $0.166$ & $0.175$  &	$-$ \\
            \textbf{\textsc{Oat}~\cite{marion2021structured} (w/ gold seed entities)} 
            & $0.250$ & $0.260$  &	$-$ \\
            \textbf{\textsc{Conquer}~\cite{kaiser2021reinforcement}} 
            & $0.240$ & $0.279$  &	$0.329$ \\
            \textbf{\textsc{Praline}~\cite{kacupaj2022contrastive}} 
            & $0.294$ & $0.373$  &	$0.464$ \\
            \textbf{\textsc{Krr}~\cite{ke2022knowledge} (w/ gold seed entities)} 
            & $\boldsymbol{0.397}$ & $0.397$  &	$0.397$   \\
        \midrule
            \textbf{$\explaignn$ (KB-only)}  & $0.330^\ddag$ & $\boldsymbol{0.399}^\ddag$	&	$\boldsymbol{0.480}^{\dag\ddag}$ \\
            \textbf{$\explaignn$}   & $0.363^\ddag$ & $\boldsymbol{0.447}^{\dag\ddag}$	&	$\boldsymbol{0.546}^{\dag\ddag}$ \\
        % \midrule
        %     \textbf{$\explaignn$ (one-shot; KB-only)}   & $0.281$ & ${0.359}$	&	${0.445}$ \\
        %     \textbf{$\explaignn$ (one-shot)}   & $0.318$ & ${0.408}$	&	${0.517}$ \\
        \bottomrule
    \end{tabular}}
    \label{tab:convquestions}
    \vspace*{-0.4cm}
\end{table}

\myparagraph{SR-attention, cross-encoding and entity types are crucial}
% We conducted an ablation study, 
% We conducted an ablation study, removing different
% mechanisms proposed, such as the SR-attention,
% the entity type (in the answering iteration),
% or cross-encodings.
Table~\ref{tab:ablation} shows results ($^\drop$ indicates significant performance drop) of our ablation study.
We found that each of these mechanisms
helps the pipeline to improve QA performance.
The most decisive factor is the
SR-attention (Sec.~\ref{sec:mp}), which ensures that only the question-relevant
information is spread within the local neighborhoods:
without this component, the performance drops substantially (P@1 from $0.442$ to $0.062$).
Similarly, the cross-encodings
(Sec.~\ref{sec:encoder})
initialize the nodes with question-relevant encodings.
Also notable is the crucial role of entity types (Sec.~\ref{sec:encoder}), that help suppress irrelevant answer candidates with mismatched answer types.
% Also the entity types helped the model to soft 
% The hallucination prevention (Sec.~\ref{sec:qu}) does not significantly improve
% the performance, but can still help to avoid simple mistakes.
% \myparagraph{Anecdotal examples}

\myparagraph{Error analysis}
We identified three key sources of error:
(i) the answer is not present in the initial graph ($53.9\%$ of error cases),
which can be mitigated by improving the QU and ER stages of the pipeline,
% (e.g. the ER might miss important evidences),
(ii) the answer is dropped when shrinking the graph ($8.1\%$),
and (iii) the answer is present in the final graph
but not identified as the correct answer ($38.0\%$).
The graph shrinking procedure is responsible for only a few errors ($2.2\%$ in the first iteration, $5.9\%$ in second),
demonstrating the viability of our iterative approach.
% To improve the final answer prediction,

\begin{table} [t] 
    \caption{Ablation study of the \explaignn pipeline.}
    %  (on the dev set)
    \vspace*{-0.4cm}
    \newcolumntype{G}{>{\columncolor [gray] {0.90}}c}
    \newcolumntype{H}{>{\setbox0=\hbox\bgroup}c<{\egroup}@{}}

    % \newcommand{\drop}{*}
    % \resizebox*{\columnwidth}{!}{
    	\begin{tabular}{l G G G}
    		\toprule
                \textbf{Method $\downarrow$}          & \textbf{P@1}  & \textbf{MRR}  & \textbf{Hit@5} \\ %  & \textbf{Runtime}\\
            \midrule
        	    \textbf{$\explaignn$}                 & $\boldsymbol{0.442}$    & $\boldsymbol{0.505}$  & $\boldsymbol{0.589}$  \\ % & $-$\,ms   \\
            \midrule
            %
                % \textbf{w/o alternating encoder}	& $-$    & $-$  & $-$  & $-$   \\
                % \textbf{w/o structure}	            & $-$    & $-$  & $-$  & $-$   \\
        	    \textbf{w/o} SR-attention	        & $0.062^{\drop}$    & $0.137^{\drop}$  & $0.178^{\drop}$ \\ % & $615$\,ms   \\
        	    \textbf{w/o} cross-encoder	        & $0.352^{\drop}$    & $0.431^{\drop}$  & $0.529^{\drop}$ \\ % & $586$\,ms   \\
                \textbf{w/o} entity type	        & $0.420^{\drop}$    & $0.492^{\drop}$  & $0.584$ \\ % & $-$\,ms   \\
                % \textbf{w/o} multi-task learning	& $0.420$*    & $0.486$*  & $0.576$  & $632$\,ms   \\                
             \midrule
                \textbf{w/o} hallucination prevention	& $0.436$    & $0.502$  & $0.588$ \\ %  & $-$\,ms   \\
                \textbf{w/o} use of SR slots in retrieval	  & $0.427^{\drop}$    & $0.493^{\drop}$  & $0.578^{\drop}$ \\ % & $-$\,ms   \\
    		\bottomrule
    	\end{tabular}
     % }
    \label{tab:ablation}
    \vspace*{-0.3cm}
\end{table}

%% file: sections/06-explainability.tex
% \clearpage
% \newpage

% \begin{figure} [t]
%      \includegraphics[width=0.8\columnwidth]{images/2023-explaignn-study-explaignnation.pdf}     
%      \vspace*{-0.4cm}
%      \caption{A representative example from our user study, illustrating a typical output of \explaignn.}
%      \label{fig:explaignnation}    
%      % \vspace*{-0.45cm}
% \end{figure}

% \begin{table*} [t] 
%     \resizebox*{\textwidth}{!}{
%         \begin{tabular}{ c | c | c }
%             \includegraphics[width=0.33\textwidth]{images/2023-explaignn-study-explaignnation.pdf}    & \includegraphics[width=0.33\textwidth]{images/2023-explaignn-study-explaignnation.pdf}     & \includegraphics[width=0.33\textwidth]{images/2023-explaignn-study-explaignnation.pdf}     \\
%             \midrule
%             User input a    & User input b      & User input c \\
%         \end{tabular}
%     }
%     \label{fig:explaignnation}   
% \end{table*}

\begin{table*} [t] 
    \resizebox*{\textwidth}{!}{
        \begin{tabular}{ l c | l c }
                \multicolumn{2}{c|}{\includegraphics[width=0.45\textwidth]{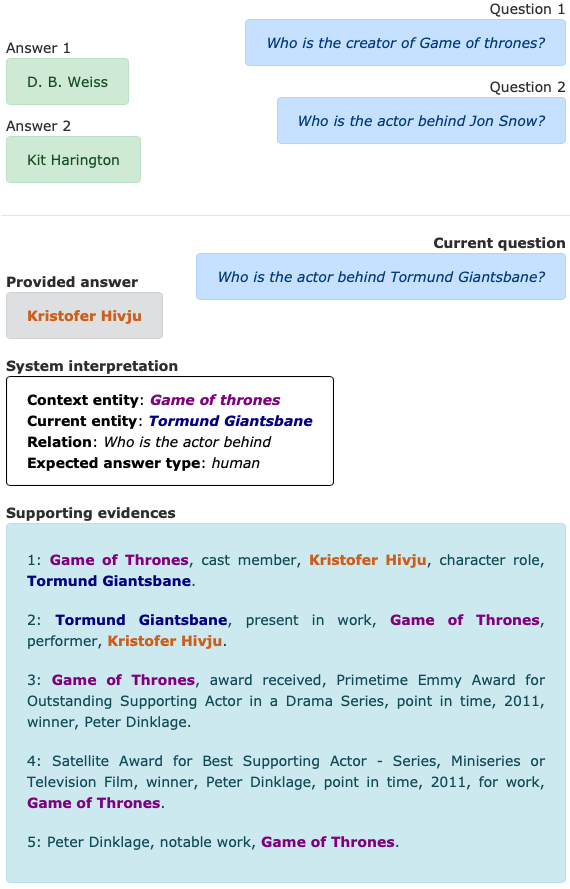}}  & \multicolumn{2}{c}{\includegraphics[width=0.5\textwidth]{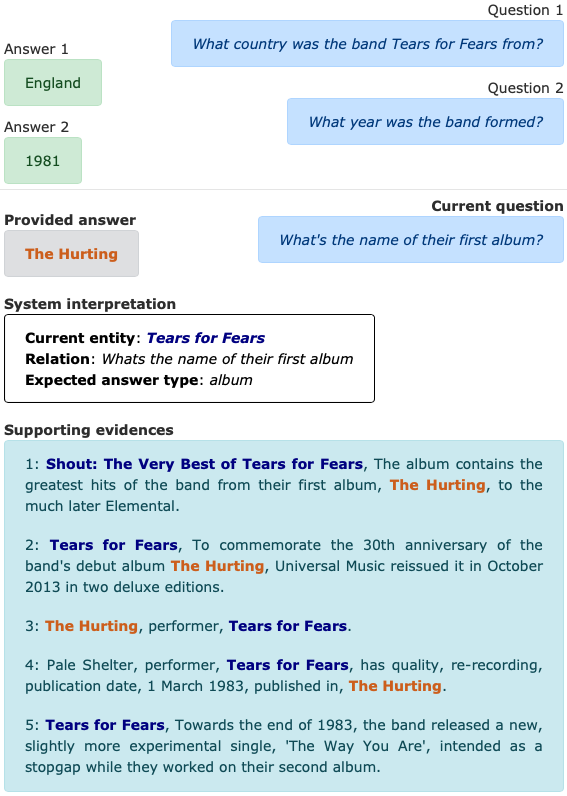}}   \\
            \midrule
                \textbf{Answer correctness} (Correctness of the provided answer) &  \cmark  &  \textbf{Answer correctness} (Correctness of the provided answer) &  \cmark    \\
                \textbf{Predicted correctness} (User assessment of the answer correctness) &  \cmark  &  \textbf{Predicted correctness} (User assessment of the answer correctness) &  \cmark    \\
                \textbf{User correctness} (Correctness of the user assessment) &  \cmark  &  \textbf{User correctness} (Correctness of the user assessment) &  \cmark    \\
            \midrule
                \textbf{User certainty} (User certainty about her assessment) &  \cmark  &  \textbf{User certainty} (User certainty about her assessment) &  \cmark    \\
        \end{tabular}
    }
    \vspace{0.2cm}
    \caption{Representative examples from our user study, illustrating typical outputs of \explaignn. In both cases, \explaignn obtained the \textit{correct} answer, and the users were \textit{certain} about their decision.}
    \label{fig:explaignnation-1}
    \vspace{-0.2cm}
\end{table*}

\begin{table*} [t] 
    \resizebox*{\textwidth}{!}{
        \begin{tabular}{ l c | l c }
                \multicolumn{2}{c|}{\includegraphics[width=0.4\textwidth]{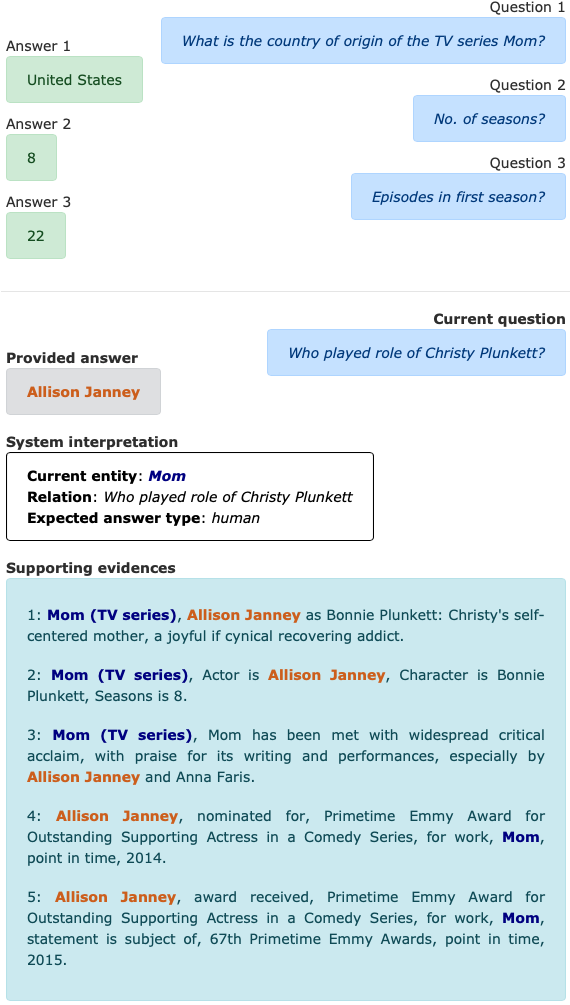}}  & \multicolumn{2}{c}{\includegraphics[width=0.5\textwidth]{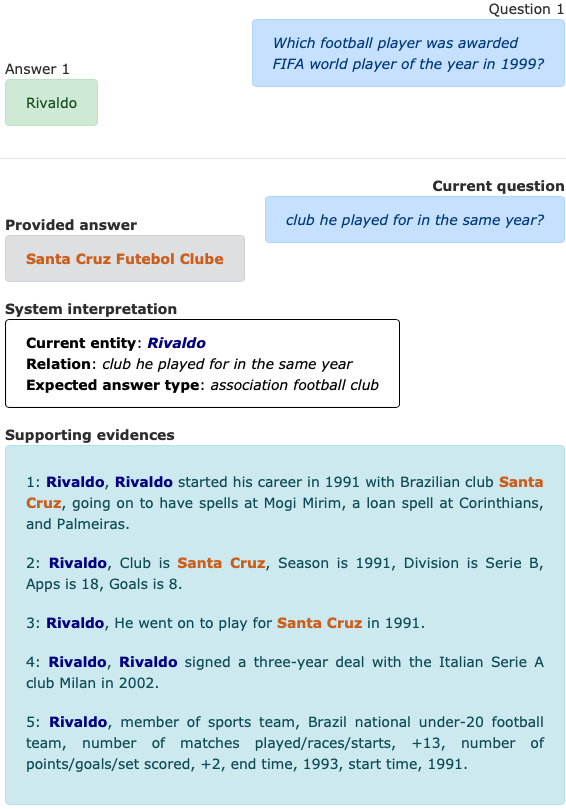}}   \\
            \midrule
                \textbf{Answer correctness} (Correctness of the provided answer) &  \xmark  &  \textbf{Answer correctness} (Correctness of the provided answer) &  \xmark    \\
                \textbf{Predicted correctness} (User assessment of the answer correctness) &  \xmark  &  \textbf{Predicted correctness} (User assessment of the answer correctness) &  \xmark    \\
                \textbf{User correctness} (Correctness of the user assessment) &  \cmark  &  \textbf{User correctness} (Correctness of the user assessment) &  \cmark    \\
            \midrule
                \textbf{User certainty} (User certainty about her assessment) &  \cmark  &  \textbf{User certainty} (User certainty about her assessment) &  \xmark    \\
        \end{tabular}
    }
    \vspace{0.2cm}
    \caption{Representative examples from our user study, illustrating cases in which \explaignn obtained an \textit{incorrect} answer. In the first case, the user could \textit{certainly} tell that the answer is incorrect from the provided evidences. In the second case, the provided evidences were not sufficient for deciding the correctness of the answer (user was \textit{uncertain}).}
    \label{fig:explaignnation-2}
    \vspace*{-0.4cm}
\end{table*}

% !TEX root = ../2023-sigir-fp-explaignn.tex
\section{User study on explainability}
\label{sec:explain}
% discuss different impacts and aspects of explainability

% In the \explaignn pipeline, we aim to derive intermediate outputs
% to enhance the pipeline's explainability for end users.
% % that can be presented to end users 
% However, what appears explainable to the authors/creators of a pipeline,
% might not easily be understood by non-expert users.

We evaluate the % investigate the
explainability of our pipeline
by a 
% dedicated
user study on Amazon Mechanical Turk (AMT).
% Our goal is to keep the user study layout general
The use case scrutinized here
% that we have in mind,
is that a user
has an information need, obtains the answer predicted by the system,
and is unsure whether to trust the provided answer.
Thus, the key objective of the explanations in this work
is to help the user decide whether to trust the answer. % or not.
If the user is able to % take
make
this decision
easily,
the explanations serve their purpose.

% describe user study
% \subsection{User study setup}

\myparagraph{User study design}
% We propose the following user study design:
For a predicted answer to a conversational question,
the user is given the conversation history, the current question, the SR,
and the explaining evidences.
% , and the predicted answer by the pipeline.
Examples of the input presented to the user
are shown in Fig.~\ref{fig:explaignnation-1} and Fig.~\ref{fig:explaignnation-2}
% (with a variant 
% of\explaignn
(we used five explanatory evidences).
The user then has to decide whether the provided answer is correct. % or not.
We randomly sample 
% a set of instances (
$1{,}200$ instances
% in this case),
on which we measure user accuracy.
% The crux of the user study is
% that we collect instances such that
One of our main considerations during sampling was that
one half ($600$) was correctly answered
and the other half ($600$) incorrectly answered by \explaignn. % system,
% as
% can be % understood
% inferred from the gold answers in the benchmark.

\myparagraph{User study interface}
For each instance, we then ask Turkers the following questions:
(i) \phrase{Do you think that the provided answer is correct?} (\textbf{User correctness}),
(ii) \phrase{Are you certain about your decision?} (\textbf{User certainty}), and
(iii) \phrase{Why are you certain about the decision?} or \phrase{Why are you uncertain about the decision?}.
The first two questions can be answered by either \phrase{Yes}, or \phrase{No}.
Depending on the answer to the second question,
the third question asks for reasons for their (un)certainty.
The user can select multiple provided options.
If the user is certain,
these options are either 
\textit{good explanation}, \textit{prior knowledge}, and \textit{common sense}.
If the user is uncertain,
then she must choose between \textit{bad explanation} or \textit{question/conversation unclear}.
% if the user is uncertain
% (on a high-level; the provided options are made more explicit in the actual interface).
The idea is to remove confounding cases in which users
% are certain or uncertain
make a decision
regardless of the provided explanation,
since we cannot infer much from these.
Note that Turkers were not allowed to access external sources like web search. %,
% and rely on the provided information only.
% - answer correct
% - certain/uncertain
% - reason for (un)certainty

\myparagraph{Quality control}
% The study is conducted on AMT.
For quality control, we restrict the participation in our user study to
Master Turkers with an approval rate of $\geq$95\%.
Further, we added honeypot questions, for which the answer and provided
information are clearly irrelevant with respect to the question (domain and answer type mismatch),
We excluded submissions from workers who gave incorrect responses to the honeypots. %  (i.e. claimed correctness).
% Filtering instances for which the user is (un)certain irrespective
% of the provided explanation also improves the quality of the collected data.
% - honeypot
% - applying filters

%%% NOT APPLICABLE, since no baseline used
% \myparagraph{Bias prevention}
% - positional bias
% - data bias

%%$$ DROPPED FOR SPACE: not that important, and can be inferred from Fig~\ref{fig:explaignnation}
% \myparagraph{Deriving explanations}
% - we make the SR accessible
% - we rank evidences
% - we mark entity mentions in evidences, matching with question entities
% - we mark answer mentions in evidences
% - baseline

\begin{table} 
    \caption{Confusion matrix for the user study, showing the probabilities of user correctness against user certainty.}
    \vspace*{-0.4cm}
    \newcolumntype{G}{>{\columncolor [gray] {0.90}}c}
    % \resizebox*{0.8\columnwidth}{!}{
        \begin{tabular}{l | c c | c} 
                & \textbf{User correct} &  \textbf{User incorrect} \\
                \hline
                    \textbf{User certain}       & {$0.632$} & {$0.166$} & $0.798$\\
                    \textbf{User uncertain}     & {$0.129$} & {$0.073$} & $0.202$\\
                \hline
                                                & $0.761$ & $0.239$ & \\
               
        \end{tabular}
        \label{tab:user-study}
    % }
    \vspace*{-0.45cm}
\end{table}

% is the pipeline explainable?
\myparagraph{\explaignn provides explainable answers}
% - how many instances collected
% - how many instances left after applying filters
% Results of the user study
Findings are presented as a confusion matrix in Table~\ref{tab:user-study} (values computed after removing confounding assessments). %  as in cases of answer known/input unclear).
% After pruning out the irrelevant assessments,
% in which users already knew the answer,
% or found the question (or conversation) confusing,
In the $771$ of the $1{,}200$ observations that remain,
% were left.
we found that the user 
can accurately decide the correctness of the system answer ($76.1$\%)
and is certain about their assessment ($79.8$\%)
most of the time.
This proves that our % the provided
explanations are indeed
% comprehensible by
useful to end users.
If the user is certain about their assessment, then we observed that the accuracy in deciding the correctness was higher:
P(User\,correct $|$ User\,certain) = $0.792$.

\myparagraph{Anecdotal examples}
Fig.~\ref{fig:explaignnation-1} and Fig.~\ref{fig:explaignnation-2}
show example outputs of \explaignn,
and the corresponding results from the user study.
For the examples in Fig.~\ref{fig:explaignnation-1} the \explaignn answer was \textit{correct}.
Based on the explanation (system interpretation and supporting evidences)
the users could \textit{certainly} tell the correctness of the answer.
For the examples in Fig.~\ref{fig:explaignnation-2}, the answer provided by \explaignn was \textit{incorrect}.
In the first case, the user was able to identify this incorrectness \textit{with certainty} from the explanation
(the first and second supporting evidences mention that Allison Janney played Bonnie Plunkett, and not Christy Plunkett).
The user could then try to reformulate~\cite{kaiser2021reinforcement} the question to obtain the correct answer.
In the second case, the provided information was not sufficient, since there is no information on the club Rivaldo played for in 1999. So the user was \textit{uncertain} about the answer correctness.
Note that even in this scenario the end user would understand that the provided answer needs to be verified.
This is in contrast to incorrect answers generated by large language models, which often look perfect on surface,
giving the user a false sense of correctness.

The guidelines, code, and the results for the user study are publicly available\footnote{\url{https://qa.mpi-inf.mpg.de/explaignn/explaignn_user_study.zip}}.

% Note that the annotation process is highly subjective:
% users may have different backgrounds, and diverse prior knowledge
% on a given domain or entity. Further, personal preferences can also play
% a role.
% Hence, measuring inter annotator agreement (IAA) is not applicable here.

% \subsection{Discussion}
% \label{sec:discussion}
% <ADD DISCUSSION ON DEVELOPER PERSONA>\\
% - will benefit from iterative reductions
% - can better investigate failure cases, fine-tune appropriate components
% A developer could, for example, analyze the graph properties within the iteration in which the gold answer is pruned, and leverage heuristic algorithms working on top of the scores produced by the GNN. A concrete example would be to enforce connectivity among evidences in the pruned graphs.

%% file: sections/07-related-work.tex
% !TEX root = ../2023-sigir-fp-explaignn.tex
\vspace{0.1cm}
\section{Related work}
\label{sec:related}

% related work on ConvQA
\myparagraph{Conversational question answering}
%%% mention:
% MMCoQA: Conversational Question Answering over Text, Tables, and Images, https://aclanthology.org/2022.acl-long.290.pdf
There has been extensive research on ConvQA~\cite{saharoy2022question, reddy2019coqa} in recent years,
% Research on ConvQA
which can largely be divided into
methods using a KB~\cite{ke2022knowledge, marion2021structured, lan2021modeling},
methods using a text corpus~\cite{qiu2021reinforced, qu2019bert, qu2020open},
and methods integrating tables~\cite{mueller2019answering,iyyer2017search}.

In \textbf{ConvQA over KBs},
% methods operating
the (potentially completed) question
is often mapped
to logical forms
that are run over the KB to obtain the answer~\cite{guo2018dialog, marion2021structured, ke2022knowledge, shen2019multi, perez2023semantic}.
A different type of approach
is to search for the answer in the local neighborhoods of an ongoing
context graph, which captures the relevant entities of the conversation~\cite{christmann2019look, lan2021modeling, kaiser2021reinforcement}.
Early \textbf{ConvQA systems over textual sources} assumed the relevant information
(i.e. text passage or document) to be given~\cite{reddy2019coqa, choi2018quac, huang2018flowqa},
and modeled the problem as a machine reading comprehension (MRC) task~\cite{rajpurkar2016squad}.
This
% strong
assumption was challenged in~\cite{qu2020open},
% and the authors 
which proposed \textsc{ORConvQA}
% , an end-to-end pipeline
introducing a retrieval stage.
Recent works follow similar ideas, and mostly rely on \textit{question rewriting}~\cite{raposo2022question, vakulenko2021question} or \textit{question resolution}~\cite{voskarides2020query},
% to target the challenges of the conversational setting,
and then employ a MRC model.
In related work on \textbf{ConvQA over tables},
% ~\cite{mueller2019answering,iyyer2017search},
the answer is either derived via logical forms~\cite{iyyer2017search},
or via pointer-networks operating on graph-encodings of the tables~\cite{mueller2019answering}.
% The assumption that the relevant table is given
% has not yet been challenged in this line of work,
% to the best of our knowledge.

All of these methods rely on a single information source for answering questions,
inherently limiting their answer coverage.
Recently, there has been preliminary work on ConvQA using a mixture of the
sources~\cite{christmann2022conversational, deng2022pacific}.
The method proposed in~\cite{deng2022pacific}
appends incoming questions to the conversational history,
and then generates a program to derive the answer from
a table using a sequence-to-sequence model.
In~\cite{christmann2022conversational}, evidences from
heterogeneous sources
are concatenated and fed into a sequence-to-sequence model
to generate the answer.
Both methods heavily rely on sequence-to-sequence models
% that require the input to be plain token sequences.
% Further,
where the generated outputs are not explainable,
and may even be % \textit{``hallucinated''} .
hallucinated.

% related work on heterogeneous QA
\myparagraph{QA over heterogeneous sources}
% cite some notable works on 2 sources: kg+text, text+table, ... 
In addition to work on ConvQA over heterogeneous sources,
there is a long line of work on answering \textit{one-off questions} using such mixtures of sources~\cite{ferrucci2012introduction,xu2016hybrid,xu2016question,savenkov2016knowledge,sun2018open,sun2019pullnet,chen2021open,pramanik2021uniqorn}.
More recently, UniK-QA~\cite{oguz2021unikqa} proposed the verbalization of evidences, and then applied FiD~\cite{izacard2021leveraging} for the answering task.
UDT-QA~\cite{ma2021open} improved over UniK-QA by implementing
more sophisticated mechanisms for evidence verbalization, and use T5~\cite{raffel2020exploring}
% as the seq2seq % sequence-to-sequence
% model
to generate the answer.
Similarly, Shen et al.~\cite{shen2022product} propose a dataset and method for answering questions on products from heterogeneous sources, leveraging BART~\cite{lewis2020bart}.
% as sequence-to-sequence model.
These approaches, being
% based on sequence-to-sequence models,
sequence-to-sequence models at their core,
face similar problems as mentioned before. % outlined above.
HeteroQA~\cite{gao2022heteroqa} explore heterogeneity in the context of community QA
% question answering (CQA)
where retrieval sources could be posts, comments, or even other questions. % and others.
% A notable aspect of this work was the use of question-relevant attention in GNNs, that is similar to our SR-attention but

% related work on explainability in QA / GNNs
\myparagraph{Explainable QA}
Existing work is mostly on single-turn methods
operating over a single source, %~\cite{yang2018hotpotqa,jia21complex,nishida2019answering},
with template~\cite{abujabal2017quint,shekarpour2020qa2explanation} and graph-based derivation sequences~\cite{lu2019answering,jia21complex,pramanik2021uniqorn} as mechanisms for ensuring explainability.
% Obtaining logical forms
% % (e.g. SPARQL queries)
% is a common explanation strategy for QA approaches using KBs~\cite{abujabal2017quint}.
Works on text-QA provide end users with actual passages and reasoning paths used for answering~\cite{nishida2019answering,yang2018hotpotqa}.
% Showing reasoning paths
% In MKHAN~\cite{zhang2019multi},
% % model-internal
% reasoning paths are used for explaining answers.
% of a medical QA system.
% In~\cite{sydorova2019interpretable}, the authors compare several general-purpose
% explanation methods for an existing QA system.
Post-hoc explainability for QA over KBs and text is investigated in~\cite{sydorova2019interpretable}. %, along with a 
These methods cannot be easily adapted to
a conversational setting with incomplete follow-up questions.
% , or provide explanations for heterogeneous sources.

\myparagraph{Explainability in GNNs}
Explainability for GNNs is an active field of research~\cite{yuan2022explainability},
devising general techniques
to identify important features for the nodes in the graphs,
or provide model-level explanations.
% or perturbate the input graphs to iden
Such approaches are mostly designed for developers,
% enhancing the explainability of developers,
% making system outputs understandable to 
and therefore hardly applicable in our scenario. % with end users.
% The most relevant technique for our scenario is
% input perturbation: nodes (or edges) are dropped from the graph,
% running the inference on a subset of nodes to identify the relevance
% of the dropped nodes for the prediction.
% A problem of applying input perturbation to our scenario (i.e. our heterogeneous graph),
% is that one cannot discriminate between the effect on the structure
% and the content, when dropping nodes.
% Further, such methods are computationally expensive, and therefore
% not applicable in ConvQA, as systems are expected to
% % provide answer and explanation 
% respond
% within hundreds of milliseconds.
Through our iterative model, 
we propose a QA-specific method for
deriving explanations
of GNN predictions
that can be understood by \textit{average web users}.
% we take one of the first steps
% in this direction, and show that end users can comprehend the
% derived explanations.

% cite paper: Iterative Deep Graph Learning for Graph Neural Networks: Better and Robust Node Embeddings

%% file: sections/08-conclusion.tex
% !TEX root = ../2023-sigir-fp-explaignn.tex
\vspace*{0.3cm}
\section{Conclusion}
\label{sec:conclusion}

There are three main takeaways for a broader audience from this work.
First, at a time when large language models (LLMs) like ChatGPT are used as a one-stop shop for most NLP tasks including ConvQA, our method \explaignn stands out by providing traceable provenance of its answer predictions.
Next, explainability for graph neural networks is an unsolved concern: we propose an iterative model that sequentially reduces the graph size as a medium for offering causal insights into the prediction process.
Finally, for several systems in IR and NLP, performance, efficiency, and explainability are seen as trade-offs. Through our
highly
configurable solution, we show that in certain use cases, it is actually possible to find configurations that lie at the sweet spots of all the three factors.
A natural future work would be to generalize these insights to other problems, like graph-based neural recommendation.